\documentstyle[aps,prl]{revtex}
\begin{document}
\draft
\author{Robert de Mello Koch and Jo\~ao P. Rodrigues,
}
\address{Department of Physics and Centre for Non Linear Studies,\\ 
University of the Witwatersrand, Wits 2050, South Africa}
\title{Systematic 1/N corrections for bosonic and fermionic vector models
without auxiliary fields. }

\maketitle
\begin{minipage}{\textwidth}
\begin{quotation}
\begin{abstract}
In this paper, colorless bilocal fields are employed to study the large $N$
limit of both fermionic and bosonic vector models. The Jacobian associated
with the change of variables from the original fields to the bilocals is
computed exactly, thereby providing an exact effective action. This effective
action is shown to reproduce the familiar perturbative expansion for the two
and four point functions. In particular, in the case of fermionic vector
models, the effective action correctly accounts for the  Fermi statistics. The
theory is also studied non-perturbatively. The stationary points of the effective
action are shown to provide the usual large $N$ gap equations. The homogeneous
equation associated with the quadratic (in the bilocals) action is simply the two
particle Bethe Salpeter equation. Finally, the leading correction in $1\over N$
is shown to be in agreement with the exact $S$ matrix of the model.
\bigskip

\noindent PACS numbers:11.10.Lm, 11.10.St, 11.15.Pg, 03.70.+k.
\bigskip
\end{abstract}
\end{quotation}
\end{minipage}

\section{Introduction}

One of the most important problems facing high energy physics today is the
construction of suitable approximation techniques that will allow an
analytic understanding of the long distance properties of non Abelian
gauge theories. The large $N$ expansion remains one of the most promising 
techniques proposed to date. 
The leading order of this approximation, the master
field, is given by the sum over planar diagrams in matrix 
models~\cite{tHooft}. It has
up to now been impossible to compute this sum in closed form, except for
simple systems~\cite{BIPZ}. 
For this reason, not even the leading term
in the expansion, in four dimensions, has been computed. In principle,
if the leading term could be obtained, one would set up a perturbation theory
around asymptotic states consisting of "mesons" and "hadrons" - colorless
bound states of the quanta of the matter fields appearing in the Lagrangian.
In other words, fluctuations about this master field yield the mass
spectrum of the theory~\cite{JeRo2}.

Vector models by contrast, have soluble large N limits. Indeed, the large
N limit of a number of vector models has been studied using an auxiliary
field~\cite{GN},~\cite{Gross}, which is not
(classically) dynamical. 
Formally eliminating the auxillary field, the
original model under consideration is regained. The advantage of the
auxiliary field is that it is chosen so that it contains no implicit N
dependance. Thus, integration over the original field variables yields
an effective theory in which all N dependence is fully explicit.
There are however, at least two serious objections to this approach. Firstly,
it is only viable for Lagrangians consisting of quadratic plus
quartic terms. Secondly, the local auxillary field is a
poor substitute for the composite (bilocal) meson field. 

Jevicki and Levine~\cite{JeLe} have constructed the master fields 
for a large number of
bosonic vector models, working directly with 
equal time bilocal meson fields, thereby
overcoming the difficulties mentioned above. They rewrite the Hamiltonian in
terms of 
these bilocal fields, obtaining 
an effective theory of mesons in
which all N dependance is fully explicit. 
This change of variables
induces a non trivial Jacobian~\cite{JS}, 
which is obtained by imposing hermiticity
on a suitably scaled Hamiltonian. This approach is not easily
implemented for fermionic vector models: 
it is a well known fact that
the Dirac hamiltonian describes 
constrained dynamics. These
constraints obscure the Hamiltonian 
approach. For this reason, the large N
limit of fermionic vector models was 
originally studied using a pseudospin
formalism~\cite{JePa}. Although this equal time approach 
does provide a possible starting point for
a systematic expansion in 1/N, the presence of the pseudospin
constraints obscures the relationship between the pseudospin
variables and quantities of interest. 
The possibility of reexpressing a fermionic vector theory in terms of
unequal time 
bilocals has been recently considered by Cavichi et al.~\cite{Ca} 
who obtained
the leading large $N$ form of the effective action directly
in terms of these bilocals, and were therefore able to 
reproduce t'Hooft's $QCD_{2}$ equation~\cite{tHooft2} for the
meson spectrum.

In matrix models,
higher order corrections in $1\over N$ have been systematically 
computed in the case of 
the c=1 matrix model string theory~\cite{DJR} directly in terms
of invariant variables. 
These corrections have been shown to be in
complete agreement with the exact solution of the model. In this case 
the effective propagator is simply that of a massless scalar
and it also turns out that the effective Hamiltonian is cubic.
The invariant variables of the matrix model correspond to the bilocals in 
vector models, so the question that we address in this article is
whether systematic
higher order corrections in ${1\over N}$ can be computed directly in
terms of the bilocal variables for vector models.

Higher order calculations have been performed for a Gross
Neveu model with scalar-scalar interactions, using the auxiliary field
approach~\cite{Scho}, or directly in terms of the original 
fermionic fields~\cite{ScatAmp} as a check of the $1\over N$ 
expansion, given that the exact $S$ 
matrix of the model is known~\cite{Zam}.

In this paper, we develop the 
necessary formalism needed to systematically compute
higher order corrections in $1\over N$,
by obtaining an exact effective action
in terms of unequal time bilocal fields.
These bilocal fields are time ordered 
and we use them to study both
bosonic and fermionic vector models. 

We obtain the exact effective action 
by explicitly calculating the full Jacobian 
associated with the change of variables from the original 
fields to the time ordered bilocals. This is achieved by
requiring that the Schwinger
Dyson equations derived directly in terms of the bilocal variables agree
with the equations derived in terms of the original 
fields~\cite{JeRo}. This provides
a unified approach for both bosonic and
fermionic models. The bosonic Jacobian agrees with the result obtained from
collective field theory~\cite{JS}. The result for the
fermionic Jacobian is new. It is remarkable that within the context of
a path integral quantization such a Jacobian exists and is exactly 
computable.

This effective action is nonlocal in time but in the context
of the functional integral quantization it generates all the 
"colorless" correlators of the quantum theory. 
For fermionic theories, since the effective action is exact 
and also valid for $N=1$, it provides an exact rewriting 
of the original theory (although for $N=1$ of course there is
no small expansion parameter). We therefore expect this approach
to have interesting applications in
condensed matter type fermionic systems. 
The formalism is also likely to be particularly relevant
for problems in which the spectrum is expected to consist only of
singlets under the global invariance of the theory, since
it then directly describes the dynamics of the physical degrees 
of freedom. Finally we wish to emphasize that this formalism
provides a (nonlocal) bosonization valid for arbitrary dimensions.
Although many of these aspects are now under study the main purpose
of this article is to obtain the effective action and confirm
its validity perturbatively.

So, perturbatively in both $1\over N$ and the
coupling constant, we consider the linear sigma model and the Gross
Neveu model (with scalar-scalar interactions) and verify that 
typical correlators are in agreement with their Feynman expansions,
and in particular verify that the statistics is 
properly taken into account.
This provides a strict test of the validity of our effective action.
We go on to describe how to develop a systematic expansion of
the effective action and explicitly establish
$1\over N$ as the perturbative parameter of the expansion. 

The paper is organized as follows: in section $II$ the 
effective action, connection to the collective field theory
and check of the perturbative expansion in both $1\over N$ 
and the coupling constant is carried out. In section $III$
we do similarily for fermionic vector models. 
In section $IV$ we make contact with the (nonperturbative) analysis
of the spectrum for Gross Neveu and 
Nambu-Jona-Lasinio type model by obtaining the propagator
of our effective field theory. We show further that the
homogeneous equation for the quadratic fluctuations of our
effective field theory is a Bethe-Salpeter equation for these
models. In section $V$ we obtain all diagrams contributing to
the $1\over N$ correction to two particle scattering from
the effective field theory and show that they are in precise agreement 
with the diagrams considered in previous comparisons of the
$1\over N$ expansion with the exact $S$ matrix~\cite{ScatAmp}.

\section{Bosonic vector models and the relationship to the collective
field theory.}

\subsection{Schwinger Dyson equations and Jacobian.}

In this section we will consider $U(N)$ invariant bosonic vector models.
By this we mean any theory of $N$ complex scalar fields $\phi^{a}(x)$,
$a=1,2,...N$ ($x$ is a $d$ dimensional parameter) with an action $S$
that is invariant under the global symmetry 
$\phi^{a}\to\phi'^{a}=U^{ab}\phi^{b}$,
$\phi^{*a}\to\phi'^{*a}=U^{ab}\phi^{*b}$ with $U^{ab}$ an arbitrary
$U(N)$ element. Furthermore we assume that all coupling constants of
the theory have been rescaled appropriately so as to yield a systematic
$1/N$ expansion. This is equivalent to the statement that a rescaling of the
fields exists under which $S\to NS$. A specific example that will be
examined in detail later in this section is given by the action

\begin{equation}
S=\int d^{d}x(\partial_{\mu}\phi^{*a}\partial^{\mu}\phi^{a}
-m^{2}\phi^{*a}\phi^{a}-{g^{2}\over 8N}(\phi^{*a}\phi^{a})^{2}),
\label{1}
\end{equation}

\noindent although our discussion applies to 
arbitrary invariant actions $S$ as 
described above. These include (non local) effective actions 
resulting from the explicit integration of intermediate
degrees of freedom such as $QCD_{2}$. This will be the object of
further study in another communication~\cite{us}.

We will assume further that one is only interested in time ordered product
expectation values of invariant operators (generically referred to as
invariant correlators)

\begin{equation}
<F[\sigma ]>\equiv <\prod^{m}_{i=1}\sigma (x_{i},y_{i})>=
{\int D\phi^{*} D\phi e^{iS}\prod^{m}_{i=1}\sigma (x_{i},y_{i})
\over\int D\phi^{*} D\phi e^{iS}}
\label{2}
\end{equation}

\noindent with

\begin{equation}
\sigma (x,y)=\phi^{*a}(x)\phi^{a}(y)
\label{3}
\end{equation}

\noindent It is straightforward to obtain a set 
of Schwinger-Dyson equations
for the invariant correlators. These follow from the identity

\begin{equation}
0=\int D\phi^{*} D\phi {\delta\over\delta\phi^{a}(x)}
\Big[ \phi^{a}(y) F[\sigma ] e^{iS} \Big]
\label{4}
\end{equation}

\noindent yielding

\begin{equation}
<N\delta^{d} (x-y)F[\sigma ]>+<\phi^{a}(y){\delta F\over\delta
\phi^{a}(x)}>+i<\phi^{a}(y){\delta S\over \delta \phi^{a}(x)}
F[\sigma ]>=0
\label{5}
\end{equation}

\noindent The above set of equations involve only 
invariant correlators as
it will be shown in the following. 

The philosophy that we wish to adopt here is that there is 
a Jacobian $J$ associated with the change of variables from the original
variables $\phi^{a}$ to the invariant variables~(\ref{3}) inside the 
functional
integral that correctly yields all time ordered product expectation 
values of invariant operators~\cite{JS}.
In other words, 

\begin{equation}
\int D\phi^{*}D\phi F[\sigma ]e^{iS}
=\int D\sigma J \, F[\sigma ] e^{iS}
\label{6}
\end{equation}

We now follow the procedure 
described in reference~\cite{JeRo} 
to obtain a differential equation for the Jacobian from the
loop equations~(\ref{5}). This results from the identity:

\begin{equation}
0=\int D\sigma \int d^{d}z {\delta\over\delta\sigma (z,x)}
(\sigma (z,y) J F[\sigma ]e^{iS}) 
\label{7}
\end{equation}

\noindent which implies\footnote{We use $L^{d}$ 
to denote the volume of
the system we are studying, i.e. $L^{d}=\int d^{d}x$.}

\begin{equation}
<\delta^{d} (x-y)\delta^{d} (0)L^{d}F[\sigma ]+\int d^{d}z\sigma (z,y)
{\delta ln J\over\delta \sigma (z,x)}F[\sigma ]
+\int d^{d}z\sigma (z,y) {\delta F[\sigma ]\over\delta\sigma(z,x)}
+i\int d^{d}z\sigma (z,y) {\delta S\over\delta\sigma(z,x)}
F[\sigma ]>=0.
\label{8}
\end{equation}

We can now use the chain rule

\begin{equation}
{\delta\over\delta\phi^{a}(x)}=
\int d^{d}z\int d^{d}y {\delta\sigma(z,y)\over\delta\phi^{a}(x)}
{\delta\over\delta\sigma(z,y)}=
\int d^{d}z\phi^{*a}(z){\delta\over\delta\sigma(z,x)}
\label{9}
\end{equation}

\noindent in equation~(\ref{5}) which is then equivalently written as

\begin{equation}
<N\delta^{d} (x-y)F[\sigma ]>+
<\int d^{d}z \sigma (z,y){\delta F\over\delta
\sigma (z,x)}>
+i<\int d^{d}z\sigma(z,y){\delta S\over \delta \sigma (z,x)}
F[\sigma ]>=0.
\label{10}
\end{equation}

Requiring agreement of this last equation with~(\ref{8}) for arbitrary
$F[\sigma ]$ we obtain

\begin{equation}
\int d^{d}z \sigma(z,y)
{\delta lnJ\over\delta\sigma (z,x)}=
(N-L^{d}\delta^{d}(0))\delta^{d}(x-y).
\label{11}
\end{equation}

J is independent of the action, as it should be. The solution to
this equation is:

\begin{equation}
lnJ=(N-L^{d}\delta^{d}(0))Trln\sigma 
\label{12}
\end{equation}

\noindent where the trace is in functional space. For models with "flavor"
degrees of freedom $\phi^{a}_{\alpha}$, $\alpha =1,2...m$ the 
invariants are 

\begin{equation}
\sigma_{\alpha\beta}(x,y)=\phi^{*a}_{\alpha}(x)
\phi^{a}_{\beta}(y)
\label{13}
\end{equation}

It is straight forward to generalize the above analysis and to
show that the Jacobian is given by

\begin{equation}
lnJ=(N-mL^{d}\delta^{d}(0))tr Trln\sigma 
\label{14}
\end{equation}

\noindent where the trace is now taken in 
both functional and flavor space.
This result generalizes for an arbitrary number of dimensions
the result obtained in~\cite{Welte} for
$d=0$ bosonic vector models.

\subsection{Connection with the collective field theory}

The point of view that the large $N$ limit can be understood in terms
of a change of variables to invariant variables or subspaces was
successfully exploited by Jevicki and Sakita~\cite{JS} for a large
class of models. In the first of~\cite{JS} these
authors are able to obtain
the form of the effective Hamiltonian acting on the reduced invariant
subspace in terms of the Jacobian associated with the new inner
product measure. Remarkably~\cite{JS} the equation satisfied by the
Jacobian can be obtained by the simple requirement that the effective
Hamiltonian must be explicitly \footnote{With 
respect to the trivial measure.} Hermitian and is given by

\begin{equation}
\sum_{C'}\Omega (C,C'){\delta lnJ\over\delta\phi (C')}
=\omega(C)-\sum_{C'}{\delta \Omega (C,C')\over\delta\phi (C')}.
\label{15}
\end{equation}

In this equation $C$ and $C'$ index the invariant variables.
In our case

\begin{equation}
\Omega (x,y;x',y')=\int d^{d-1}z {\delta \over\delta\phi^{*a}(z)}
\sigma^{*} (x,y){\delta\over\delta\phi^{a}(z)}
\sigma (x',y')
=\delta^{d-1} (y-y')\sigma (x',x)
\label{16}
\end{equation}

\noindent and

\begin{equation}
\omega (x,y)= \int d^{d-1}z{\delta^{2} 
\over\delta\phi^{*a}(z)\delta\phi^{a}(z)}
\sigma (x,y)
= N\delta^{d-1} (x-y).
\label{17}
\end{equation}

The above equations only involve equal time correlators as it is appropriate
for a Hamiltonian approach. 

Jevicki and Sakita, in the
second of~\cite{JS} have shown that exactly the same equation must be
satisfied in a functional integral description provided the invariants
are appropriately labeled i.e. in our case if $x$ and $y$ 
are $d$ dimensional spacetime points. 
With this proviso, it is straightforward to show that
once equations~(\ref{16}) and~(\ref{17}) are used in 
equation~(\ref{15}) we reproduce the
differential equation~(\ref{11}) for the Jacobian 
derived in the previous subsection.

\subsection{Perturbative Check}

The full Jacobian~(\ref{12}) is not new; 
it has certainly been written down
in~\cite{JePa}. However in all applications that we are aware of
only the leading large $N$ term in the Jacobian has been used
to obtain the leading contribution to the free energy,
leading time ordered correlators and spectrum~\cite{JeRo2}.
It is actually not clear in what sense the 
second term in the Jacobian is 
"subleading" due to infinities appearing in it
and it is not at all 
obvious how
that it can help generate systematic $1/N$ corrections. 
It is the purpose of this subsection to show that 
this is indeed the case,
perturbatively in the coupling constant.

\subsubsection{Exact Effective Action}

Any invariant correlator can be calculated from 

\begin{equation}
<F[\sigma ]>=
{\int D\sigma Je^{iS}F[\sigma ]\over
\int D\sigma Je^{iS}}=
{\int D\sigma e^{iS_{eff}}F[\sigma ]\over
\int D\sigma e^{iS_{eff}}}
\label{18}
\end{equation}

\noindent where

\begin{equation}
S_{eff}=-ilnJ+S=-iN Trln\sigma +S+iL^{d}\delta^{d}(0)Trln\sigma
\label{19}
\end{equation}

In order to exhibit explicitly the $N$ dependence we rescale
$\sigma \to N\sigma$ under which, 
as explained in subsection A,
$S\to NS$. Throughout the rest of this section this rescaling 
is always implicit. Therefore we have:

\begin{equation}
S_{eff}=-iN Trln\sigma + NS +iL^{d}\delta^{d}(0)Trln\sigma
=NS_{0}+S_{1}.
\label{20}
\end{equation}

We see that as $N\to\infty$ the leading ("classical") configuration is
determined by 

\begin{equation}
{\delta S_{0}\over\delta\sigma}|_{\sigma^{0}}=0
\label{21}
\end{equation}

We can now perturb about $\sigma^{0}$ by letting 

\begin{equation}
\sigma (x,y)=\sigma^{0}(x,y)+{1\over\sqrt{N}}\eta (x,y)
\label{22}
\end{equation}

\noindent and expanding $S_{eff}$ as

\begin{equation}
S_{eff}=NS_{0}(\sigma^{0})+S_{1}(\sigma^{0})
+{i\over 2}B_{2}+{1\over 2}A_{2}+
\sum_{n=1}^{\infty}{1\over\sqrt{N^{n}}} \big[
-i{(-1)^{n+1}\over (n+2)}B_{n+2}+
iL^{d}\delta^{d}(0){(-1)^{n+1}\over n}B_{n}
+{1\over (n+2)!}A_{n+2}\big]
\label{23}
\end{equation}

\noindent where

\begin{equation}
A_{n}=\int dx_{1}...\int dx_{n}\int dy_{1} ... \int dy_{n}
{\delta^{n} S\over\delta\sigma (x_{1},y_{1}) .....
\delta\sigma (x_{n},y_{n})}|_{\sigma^{0}}
\eta (x_{1},y_{1}).....\eta (x_{n},y_{n})
\label{24}
\end{equation}

\noindent and \footnote{The expansion of $Trln\sigma$
in this fashion is justified by the fact that
translational invariance requires
$\sigma^{0}$ to be diagonal in
momentum space.}

\begin{equation}
B_{n}= Tr ({\sigma^{0}}^{-1}\eta )^{n}.
\label{25}
\end{equation}

We obtained an effective action with an infinite number of vertices
as it is to be expected from any loop expansion. It should however
be remembered that in order to calculate any diagram to a given
order of $1\over\sqrt{N}$ only a finite number of vertices need
to be included. Notice that the subleading term of the 
Jacobian~(\ref{12}) induces tadpole type interactions. In the case of
$c=1$ strings where the effective Hamiltonian consists of
a cubic and a tadpole interaction this tadpole interaction has
been shown to be essential for an agreement with exact 
results~\cite{DJR}. For $d=0$ vector models similar arguments
have been presented in reference~\cite{Welte}. We will see in
the following that the tadpole interaction will be essential to
obtain agreement with a perturbative Feynman analysis of the
$(\phi^{*a}\phi^{a})^{2}$ theory. 

\subsubsection{Perturbative Results for the 
$(\phi^{*a}\phi^{a})^{2}$ theory.}

We will now be considering in detail the theory defined by the
action~(\ref{1})

\begin{equation}
S=\int d^{d}x(\partial_{\mu}\phi^{*a}\partial^{\mu}\phi^{a}
-m^{2}\phi^{*a}\phi^{a}-{g^{2}\over 8N}(\phi^{*a}\phi^{a})^{2})
\nonumber
\end{equation}

We will need the following perturbative result for the two
point function:

\begin{equation}
<\phi^{a}(x)\phi^{*a}(y)>=\int {d^{d}p\over (2\pi )^{d}}
G_{2} (p)
\label{26}
\end{equation}

\noindent where

\begin{equation}
\setlength{\unitlength}{1.5mm}
\begin{picture}(100,20)
\put(5,15){$G_{2}(p)=N($}
\put(18,15){\line(1,0){8}}
\put(28,15){$-2{ig^{2}\over 8}$}
\put(36,15){\line(1,0){8}}
\put(40,16){\circle{2}}
\put(46,15){$+4({-ig^{2}\over 8})^{2}$}
\put(56,15){\line(1,0){8}}
\put(58,16){\circle{2}}
\put(62,16){\circle{2}}
\put(66,16){$+4({-ig^{2}\over 8})^{2}$}
\put(76,15){\line(1,0){8}}
\put(80,16){\circle{2}}
\put(80,18){\circle{2}}
\put(86,15){$+O(g^{6}))$}
\put(7,5){$+(-2{ig^{2}\over 8}$}
\put(17,5){\line(1,0){8}}
\put(21,6){\circle{2}}
\put(27,5){$+8({-ig^{2}\over 8})^{2}$}
\put(37,5){\line(1,0){8}}
\put(39,6){\circle{2}}
\put(43,6){\circle{2}}
\put(47,6){$+8({-ig^{2}\over 8})^{2}$}
\put(57,5){\line(1,0){8}}
\put(61,6){\circle{2}}
\put(61,8){\circle{2}}
\put(67,5){$+4({-ig^{2}\over 8})^{2}$}
\put(79,6){\line(1,0){8}}
\put(83,6){\circle{2}}
\put(87,5){$+O(g^{6})) +O({1\over N})$}
\end{picture}
\label{27}
\end{equation}

For the connected piece of the four point function

\begin{eqnarray}
\nonumber
<\phi^{*a}(x_{1})&&\phi^{a}(y_{1})\phi^{*b}(x_{2})\phi^{b}(y_{2})>
\quad =\quad\int {dp_{1}\over (2\pi )^{d}}\int {dp_{2}\over (2\pi )^{d}}
\int {dp_{3}\over (2\pi )^{d}}\int {dp_{4}\over (2\pi )^{d}}
(2\pi )^{d}\delta (p_{1}+p_{3}-p_{2}-p_{4})\times \\
&&\times e^{i(-p_{2}y_{1}-p_{4}y_{2}+p_{3}x_{2}+p_{1}x_{1})}
G_{4}^{conn}(p_{1},p_{2},p_{3},p_{4}),
\label{28}
\end{eqnarray}

\noindent where

\begin{equation}
\setlength{\unitlength}{1.5mm}
\begin{picture}(100,10)
\put(5,5){$G_{4}^{conn}(p_{1},p_{2},p_{3},p_{4})=N({-ig^{2}\over 4}$}
\put(37,9){\line(1,-1){5}}
\put(37,4){\line(1,1){5}}
\put(43,5){$-{g^{4}\over 16}$}
\put(49,7.75){\line(1,0){7.5}}
\put(49,5.25){\line(1,0){7.5}}
\put(52.5,6.5){\circle{2.5}}
\put(58.5,5){$-{g^{4}\over 16}$}
\put(64,9){\line(1,-1){5}}
\put(64,4){\line(1,1){5}}
\put(66,8){\circle{1.5}}
\put(71,5){$+O(g^{6}))+O(1).$}
\end{picture}
\label{29}
\end{equation}

In the above expressions the diagrams have the standard Feynman
interpretation (without symmetry factors and coupling constants). 
For instance

\begin{equation}
\setlength{\unitlength}{1mm}
\begin{picture}(150,10)
\put(40,5){\line(1,0){8}}
\put(44,6){\circle{2}}
\put(50,5){$\quad =\quad 
{i\over p^{2}-m^{2}}\int {d^{d}k\over (2\pi )^{d}}
{i\over k^{2}-m^{2}}{i\over p^{2}-m^{2}} $}
\end{picture}
\label{30}
\end{equation}

We assume that all diagrams have been suitably regularized.
Our aim in this section is not to discuss issues related to 
renormalization, but to confirm that our effective theory
systematically reproduces the $1/N$ expansion, perturbatively
in $g^{2}$.

\subsubsection{The Leading Order.}

With a translationally invariant ansatz

\begin{equation}
\sigma (x,y) = \int {dp\over (2\pi )^{d}} e^{ip(x-y)}\sigma (p)
\label{31}
\end{equation}

It is straightforward to verify that the solution to
equation~(\ref{21}) $\sigma^{0}(p)$ satisfies

\begin{equation}
\sigma^{0}(p)={i\over p^{2}-m^{2}-{g^{2}\over 4}\int
{d^{d}k\over (2\pi)^{d}} \sigma^{0}(k)}
\label{32}
\end{equation}

This is the familiar gap equation. Iterating this equation to
$O(g^{4})$ one obtains

\begin{equation}
\setlength{\unitlength}{1mm}
\begin{picture}(150,10)
\put(20,5){$\sigma^{0}(p)=$}
\put(37,5){\line(1,0){8}}
\put(47,5){$-2{ig^{2}\over 8}$}
\put(57,5){\line(1,0){8}}
\put(61,6){\circle{2}}
\put(65,5){$+4({-ig^{2}\over 8})^{2}$}
\put(80,5){\line(1,0){8}}
\put(82,6){\circle{2}}
\put(86,6){\circle{2}}
\put(94,6){$+4({-ig^{2}\over 8})^{2}$}
\put(110,5){\line(1,0){8}}
\put(114,6){\circle{2}}
\put(114,8){\circle{2}}
\put(122,5){$+O(g^{6})$}
\end{picture}
\label{33}
\end{equation}

\noindent in agreement with the $O(N)$ term in equation~(\ref{27}).
(We recall that in our effective theory the fields are rescaled
by factors of $N$.)

\subsubsection{Effective Field Theory Spectrum}

The leading order $\sigma^{0}$ can be used, at the level of 
quadratic fluctuations, to obtain the mass spectrum
of the theory~\cite{JeRo2}. 
Perturbing about the leading order, as has been described, we 
obtain the (leading) quadratic action 

\begin{eqnarray}                              
\nonumber
S_{2}&&={i\over 2}B_{2}+{1\over 2}A_{2} \\
&&= {1\over 2}\int dx_{1}dx_{2}dx_{3}dx_{4}\big[
{-g^{2}\over 4}\delta(x_{1}-x_{2})\delta(x_{1}-x_{3})
\delta(x_{1}-x_{4})
+i{\sigma^{0}}^{-1}(x_{1},x_{2})
{\sigma^{0}}^{-1}(x_{3},x_{4})\big]\eta(x_{4},x_{1})
\eta(x_{2},x_{3}).
\label{54}
\end{eqnarray}

After introducing a symmetric Fourier transform

\begin{equation}
\eta (x_{1},x_{2})=\int {d^{d}p_{1}\over (2\pi )^{d/2}}
\int {d^{d}p_{2}\over (2\pi )^{d/2}}e^{i(p_{1}x_{1}-p_{2}x_{2})}
\eta (p_{1},p_{2})
\end{equation}

\noindent we can write the quadratic action in momentum space

\begin{equation}
S_{2}={1\over 2}\int dp_{1}dp_{2}dp_{3}dp_{4}
\big[ {-g^{2}\over 4(2\pi )^{d}}
\delta (p_{1}-p_{2}+p_{3}-p_{4}) 
+i\delta (p_{1}-p_{2})
\delta (p_{3}-p_{4}){\sigma^{0}}^{-1}(p_{1})
{\sigma^{0}}^{-1}(p_{3})\big]\eta (p_{4},p_{1})\eta(p_{2},p_{3})
\end{equation}

This action determines the propagator $A(p_{1},p_{2},p_{3},p_{4})$
of the $\sigma$ field. Conventional arguments show that 
$A(p_{1},p_{2},p_{3},p_{4})$ satisfies the equation:

\begin{eqnarray}
\nonumber
\int dp_{1}dp_{2}&&\Big[-{g^{2}\over 4}
{\delta (k_{1}+p_{1}-k_{2}-p_{2})\over (2\pi )^{d}}
+i\delta (k_{1}-p_{2}){\sigma^{0}}^{-1}(p_{1})
\delta (p_{1}-k_{2}){\sigma^{0}}^{-1}(p_{2})\Big]\times\\
&&\times A(p_{1},p_{2},p_{3},p_{4})=i\delta (k_{2}-p_{4})
\delta (k_{1}-p_{3})
\label{55}
\end{eqnarray}

Now, the physical interpretation of  
$A(p_{1},p_{2},p_{3},p_{4})$ is as follows:
this propagator will (at most) propagate a single two particle
bound state and a composite two
particle state. 
The most general ansatz with momentum conservation
consistent with this physical picture is given by:

\begin{equation}
\setlength{\unitlength}{1mm}
\begin{picture}(150,20)
\put(0,15){$A(p_{1},p_{2},p_{3},p_{4})=\delta (p_{1}-p_{4})
\delta (p_{2}-p_{3})F(p_{1},p_{2})+
\delta (p_{1}+p_{3}-p_{2}-p_{4})
G(p_{1},p_{2},p_{3},p_{4})$}
\put(20,5){$=$}
\put(25,7){\line(1,0){4}}
\put(27,7){\circle*{1.5}}
\put(25,3){\line(1,0){4}}
\put(27,3){\circle*{1.5}}
\put(29,7){\line(1,-1){5}}
\put(29,3){\line(1,1){5}}
\put(38,5){$+$}
\put(44,7.5){\line(1,-1){5}}
\put(44,2.5){\line(1,1){5}}
\put(46.5,5){\circle*{3}}
\end{picture}
\label{56}
\end{equation}

Inserting this into (\ref{55}), we immediatly find:

\begin{eqnarray}
\nonumber
F(p_{1},p_{2})&=&\sigma^{0}(p_{1})\sigma^{0}(p_{2}) \\
G(p_{1},p_{2},p_{3},p_{4})&=&-i{g^{2}\over 4}
{1\over (2\pi )^{d}}\sigma^{0}(p_{1})
\sigma^{0}(p_{2})\sigma^{0}(p_{3})\sigma^{0}(p_{4})
-i{g^{2}\over 4}\sigma^{0}(p_{1})\sigma^{0}(p_{2})
\int {d^{d}k\over (2\pi )^{d}}G(k,p_{2}-p_{1}+k,p_{3},p_{4})
\label{60}
\end{eqnarray}

Iterating this second equation for 
$G(p_{1},p_{2},p_{3},p_{4})$, it is not hard to see that
we are reproducing the series expansion for:

\begin{equation}
G(p_{1},p_{2},p_{3},p_{4})=-i{g^{2}\over 4}
\sigma^{0}(p_{1})\sigma^{0}(p_{2})\sigma^{0}(p_{3})
\sigma^{0}(p_{4})\Big[{1\over 1+i{g^{2}\over 4}
\int {d^{d}k\over (2\pi )^{d}}
\sigma^{0}(k)\sigma^{0}(k+p_{2}-p_{1})}\Big]
\label{61}
\end{equation}

It is easy to understand this last equation in terms of more familiar
approaches to the large $N$ limit: The object
$\int \sigma\sigma$ is a bosonic bubble. Thus
$G(p_{1},p_{2},p_{3},p_{4})$ is simply a sum over chains of bubble
diagrams. The factor in square braces is in fact the propagator of
the auxiliary field usually introduced to study this model~\cite{GN}.
Notice however, that our $\sigma$ propagator
$A(p_{1},p_{2},p_{3},p_{4})$ consists of two terms. The first term,
which has no analog in the auxiliary field approach, is crucial to
obtain a systematic expansion.

The mass spectrum of the theory is determined by searching for
the poles of the propagator $A(p_{1},p_{2},p_{3},p_{4})$. 
Rather than performing a full non-perturbative analysis of the 
spectrum, we content ourselves with constructing 
$A(p_{1},p_{2},p_{3},p_{4})$ to $O(g^{4})$, by iteration.
A straightforward calculation shows that 
the connected piece of $A(p_{1},p_{2},p_{3},p_{4})$,
to $O(g^{4})$ is given by

\begin{eqnarray}
\nonumber
A^{conn}(p_{1},p_{2},p_{3},p_{4})&&=
-i{g^{2}\over 4}
{\delta (p_{1}-p_{2}+p_{3}-p_{4})\over (2\pi )^{d}}
\sigma^{0}(p_{1})\sigma^{0}(p_{2})
\sigma^{0}(p_{3})\sigma^{0}(p_{4})
+(i)^{2}{g^{4}\over 16}
{\delta (p_{1}-p_{2}+p_{3}-p_{4})\over (2\pi )^{d}} \\
&&\sigma^{0}(p_{1})\sigma^{0}(p_{2})
\sigma^{0}(p_{3})\sigma^{0}(p_{4})
\int {d^{d}k\over (2\pi )^{d}}
\sigma^{0}(k+p_{2}-p_{1})\sigma^{0}(k).
\label{57}
\end{eqnarray}

Inserting the expression for the leading order (\ref{33}), yields
complete agreement 
with the $O(N)$ term in equation (\ref{29}), as it should.

\subsubsection{Corrections to the one and two point functions
of the effective theory.}

In order to compute the $O({1\over\sqrt{N}})$ 
correction to the effective theory one point function,
we need to compute the tadpole, and cubic interaction 
vertices of the effective field theory. 
These interactions arise from
the actions

\begin{eqnarray}
\nonumber
S_{tp} &=& iL^{d}\delta^{d}(0){1\over\sqrt{N}}B_{1} \\
S_{c} &=&  -i{1\over 3\sqrt{N}}B_{3}
\label{58}
\end{eqnarray}

\noindent respectivley.
For completeness,
we also present the $1\over N$ quartic and
subleading quadratic vertices

\begin{eqnarray}
\nonumber
S_{sq} &=& -iL^{d}\delta^{d}(0){1\over 2N}B_{2} \\
S_{q} &=&  i{1\over 4N}B_{4}
\end{eqnarray}

Standard techniques yield the following Feynman rules:

\begin{equation}
\setlength{\unitlength}{1mm}
\begin{picture}(150,60)
\put(10,55){\line(0,1){5}}
\put(10,59){\circle*{2}}
\put(12,55){$p_{1},p_{2}$}
\put(35,56){$-{L^{d}\delta^{d}(0)\over\sqrt{N}}
[{\sigma^{0}}^{-1}]_{p_{2}p_{1}}$}
\put(5,43){\line(1,0){10}}
\put(10,43){\circle*{2}}
\put(0,45){$k_{1},k_{2}$}
\put(13,45){$p_{1},p_{2}$}
\put(35,44){${L^{d}\delta^{d}(0)\over 2N}[{\sigma^{0}}^{-1}]_{k_{2}p_{1}}
[{\sigma^{0}}^{-1}]_{p_{2}k_{1}}$}
\put(5,31){\line(1,0){10}}
\put(10,31){\line(0,1){5}}
\put(0,28){$k_{1},k_{2}$}
\put(13,28){$p_{1},p_{2}$}
\put(6,36){$q_{1},q_{2}$}
\put(35,32){${1\over 3!\sqrt{N}}
([{\sigma^{0}}^{-1}]_{k_{2}p_{1}}
[{\sigma^{0}}^{-1}]_{p_{2}q_{1}}[{\sigma^{0}}^{-1}]_{q_{2}k_{1}}
+[{\sigma^{0}}^{-1}]_{p_{2}k_{1}}
[{\sigma^{0}}^{-1}]_{k_{2}q_{1}}[{\sigma^{0}}^{-1}]_{q_{2}p_{1}})$}
\put(6,10){\line(1,1){8}}
\put(14,10){\line(-1,1){8}}
\put(0,6.5){$k_{1},k_{2}$}
\put(13,6.5){$p_{1},p_{2}$}
\put(0,19.5){$q_{1},q_{2}$}
\put(13,19.5){$l_{1},l_{2}$}
\put(35,21){${-1\over 4!N}([{\sigma^{0}}^{-1}]_{k_{2}p_{1}}
[{\sigma^{0}}^{-1}]_{p_{2}q_{1}}[{\sigma^{0}}^{-1}]_{q_{2}l_{1}}
[{\sigma^{0}}^{-1}]_{l_{2}k_{1}}
+[{\sigma^{0}}^{-1}]_{k_{2}p_{1}}
[{\sigma^{0}}^{-1}]_{p_{2}l_{1}}[{\sigma^{0}}^{-1}]_{l_{2}q_{1}}
[{\sigma^{0}}^{-1}]_{q_{2}k_{1}}$}
\put(35,13){$+[{\sigma^{0}}^{-1}]_{k_{2}q_{1}}
[{\sigma^{0}}^{-1}]_{q_{2}l_{1}}
[{\sigma^{0}}^{-1}]_{l_{2}p_{1}}[{\sigma^{0}}^{-1}]_{p_{2}k_{1}}
+[{\sigma^{0}}^{-1}]_{k_{2}q_{1}}
[{\sigma^{0}}^{-1}]_{q_{2}p_{1}}
[{\sigma^{0}}^{-1}]_{p_{2}l_{1}}[{\sigma^{0}}^{-1}]_{l_{2}k_{1}}$}
\put(35,5){$+[{\sigma^{0}}^{-1}]_{k_{2}l_{1}}
[{\sigma^{0}}^{-1}]_{l_{2}q_{1}}
[{\sigma^{0}}^{-1}]_{q_{2}p_{1}}[{\sigma^{0}}^{-1}]_{p_{2}k_{1}}
+[{\sigma^{0}}^{-1}]_{k_{2}l_{1}}
[{\sigma^{0}}^{-1}]_{l_{2}p_{1}}
[{\sigma^{0}}^{-1}]_{p_{2}q_{1}}[{\sigma^{0}}^{-1}]_{q_{2}k_{1}})$}
\end{picture}
\label{59}
\end{equation}

In the above, we have employed an obvious matrix notation.
Recall that $\sigma^{0}(k,p)$ is diagonal in momentum space,
so that

\begin{equation}
\nonumber
[{\sigma^{0}}^{-1}]_{kp}={\delta (k-p)\over \sigma^{0}(p)}
\end{equation}

Now, we turn to the computation 
of the $O({1\over\sqrt{N}})$ correction to the
one point function $<\sigma >$. Two processes contribute: the cubic
tadpole and the linear tadpole. Using the expression (\ref{57})
for the propagator, and the above rules for the vertices, we obtain
the following expressions for the cubic tadpole

\begin{equation}
\setlength{\unitlength}{1mm}
\begin{picture}(150,25)
\put(20,12){\line(0,1){8}}
\put(18.75,11.25){$\times$}
\put(20,22){\circle{4}}
\put(25,16){$={L^{d}\delta^{d} (0)\over\sqrt{N}}\int 
dp{\sigma^{0}}^{-1}(p)A(p,p,p_{1},p_{2})
+\delta (p_{1}-p_{2}){-ig^{2}\over4\sqrt{N}}
\sigma^{0}(p_{1})\sigma^{0}(p_{2})
\Big[ \int {d^{d}k\over (2\pi )^{d}} \sigma^{0}(k)-$}
\put(25,6){$-{ig^{2}\over 4}\int {d^{d}k\over (2\pi )^{d}}
\sigma^{0}(k)\int{d^{d}l\over (2\pi )^{d}}
\sigma^{0}(l+k-p_{1})\sigma^{0}(l)
-{ig^{2}\over 4}\int{d^{d}k\over (2\pi )^{d}}
\sigma^{0}(k)\int {d^{d}l\over (2\pi )^{d}}
(\sigma^{0}(l))^{2}\Big] $}
\end{picture}
\label{62}
\end{equation}

\noindent up to $O(g^{4})$ and for the linear tadpole

\begin{equation}
\setlength{\unitlength}{1mm}
\begin{picture}(150,15)
\put(20,2){\line(0,1){8}}
\put(18.75,1.25){$\times$}
\put(20,10){\circle*{2}}
\put(25,6){$=-{L^{d}\delta^{d} (0)\over\sqrt{N}}\int 
dp{\sigma^{0}}^{-1}(p)A(p,p,p_{1},p_{2})$}
\end{picture}
\label{63}
\end{equation}

Notice that separately both the linear tadpole and cubic tadpole 
contain momentum dependent infinities. The sum however is a well
defined quantity, with all divergences proportional to 
$L^{d}\delta^{d} (0)$ cancelling and

\begin{equation}
<\eta >= \delta (p_{1}-p_{2}){-ig^{2}\over4\sqrt{N}}
\sigma^{0}(p_{1})\sigma^{0}(p_{2})
\int {d^{d}k\over (2\pi )^{d}} \Big[\sigma^{0}(k)-
{i g^{2}\over 4}\int {d^{d}l\over (2\pi )^{d}}
\Big[(\sigma^{0}(l))^{2}\sigma^{0}(k)+
\sigma^{0}(k)\sigma^{0}(l)
\sigma^{0}(k+l-p_{1})\Big]\Big] 
\label{64}
\end{equation}

Inserting the expression  (\ref{33})
for $\sigma^{0}$, 
and recalling that our fields are rescaled by a factor of $N$,
one obtains complete agreement with the perturbative 
result (\ref{27}).
This analysis shows that the linear tadpole is essential to obtain
agreement with the Feynman perturbative analysis and to remove
momentum dependent infinities.

\section{Fermionic vector models.}

In this section we show that the approach followed in section $2$ 
for bosonic vector models also applies to fermionic vector models.
This is an important observation and this is not only because fermionic
systems have more important physical applications as we now explain. 

By and large there have been
two approaches to the large $N$ limit of vector models: in the first
one, auxiliary fields are used. We have already mentioned in the
introduction some of the shortcomings of this approach.
The other approach is based on the
collective field theory~\cite{JS} idea of changing variables to invariant
quantities. However as it has been discussed in section $2b$ the 
equation satisfied by the appropriate Jacobian ultimately
stems from a Hermiticity requirement of the Hamiltonian. 
It is not obvious how
to generalize this Hamiltonian
approach to fermionic systems, although
some partial success has been achieved in terms of pseudospin
variables~\cite{JePa}. We will not pursue this method in this article
but will show that, 
provided that one is prepared to consider time ordered product
expectation values,
the fact that the Schwinger-Dyson (loop) equations 
of the theory imply a differential equation for the Jacobian,
as it was demonstrated in section $2a$, 
is straightforwardly generalizable to fermionic systems.
Moreover our approach will be fully covariant.

\subsection{Schwinger Dyson equations and Jacobian}

We assume again that we are dealing with $U(N)$ invariant
actions i.e. actions invariant under
$\psi_{\alpha}^{a}\to{\psi'_{\alpha}}^{a}=U^{ab}\psi_{\alpha}^{b}$,
$\psi_{\alpha}^{*a}\to{\psi'_{\alpha}}^{a}=U^{ab}\psi_{\alpha}^{*b}$ 
with $U^{ab}$ an arbitrary $U(N)$ element. $\alpha$ is a Dirac
index. The invariants are now

\begin{equation}
\sigma_{\alpha\beta} (x,y)=\bar{\psi}_{\alpha}^{a}(x)\psi_{\beta}^{a}(y)
\label{34}
\end{equation}

\noindent and as before we are interested in time ordered product expectation
values of invariant opertators

\begin{equation}
<F[\sigma ]>\equiv <\prod^{m}_{i=1}\sigma_{\alpha_{i}\beta_{i}} 
(x_{i},y_{i})>=
{\int D\psi^{*} D\psi e^{iS}\prod^{m}_{i=1}
\sigma_{\alpha_{i}\beta_{i}}  (x_{i},y_{i})
\over\int D\psi^{*} D\psi e^{iS}}
\label{35}
\end{equation}

A set of Schwinger-Dyson equations for the invariants follow from
the identity

\begin{equation}
0=\int D\psi^{*} D\psi {\delta\over\delta\psi_{\beta}^{a}(x)}
\Big[ \psi_{\rho}^{a}(y) F[\sigma ] e^{iS} \Big]
\label{36}
\end{equation}

It is important to remember that the fields $\psi$ have to be
treated as Grassman variables in all manipulations to follow.
We obtain

\begin{equation}
<N\delta_{\beta\rho}\delta^{d} (x-y)F[\sigma ]>
-<\psi_{\rho}^{a}(y){\delta F\over\delta
\psi_{\beta}^{a}(x)}>
-i<\psi_{\rho}^{a}(y){\delta S\over \delta \psi_{\beta}^{a}(x)}
F[\sigma ]>=0
\label{37}
\end{equation}

Postulating the existence of a Jacobian $J$ defined by

\begin{equation}
\int D\psi^{*}D\psi F[\sigma ]e^{iS}
=\int D\sigma J \quad F[\sigma ] e^{iS}
\label{38}
\end{equation}

\noindent we consider the identity

\begin{equation}
0=\int D\sigma \int d^{d}z {\partial\over\partial
\sigma_{\alpha\beta} (z,x)}
(\sigma_{\alpha\rho} (z,y) J F[\sigma ]e^{iS}) 
\label{39}
\end{equation}

\noindent which implies

\begin{eqnarray}
\nonumber
<m\delta_{\beta\rho}&&\delta^{d} (x-y)\delta^{d} (0)L^{d}F[\sigma ]
+\int d^{d}z\sigma_{\alpha\rho} (z,y)
{\partial ln J\over\partial 
\sigma_{\alpha\beta} (z,x)}F[\sigma ]
+\int d^{d}z
\sigma_{\alpha\rho} (z,y) {\partial F[\sigma ]\over\partial
\sigma_{\alpha\beta} (z,x)} \\
&&+i\int d^{d}z
\sigma_{\alpha\rho} (z,y) {\partial S\over\partial
\sigma_{\alpha\beta}(z,x)}
F[\sigma ]>=0.
\label{40}
\end{eqnarray}

In the above equation $m$ is the dimension of the representation of the
Clifford algebra.
Using the chain rule

\begin{equation}
{\delta\over\delta\psi_{\alpha}^{a}(x)}=
\int dz\int dy {\delta\sigma_{\beta\rho}(y,z)
\over\delta\psi_{\alpha}^{a}(x)}
{\delta\over\delta\sigma_{\beta\rho}(y,z)}=
-\int dy\bar{\psi}_{\beta}^{a}(y){\delta\over
\delta\sigma_{\beta\alpha}(y,x)}
\label{41}
\end{equation}

\noindent in equation~(\ref{37}) it becomes

\begin{equation}
<N\delta_{\beta\rho}\delta^{d} (x-y)F[\sigma ]>-
<\int dz \sigma_{\alpha\rho} (z,y){\delta F\over\delta
\sigma_{\alpha\beta} (z,x)}>
-i<\int dz\sigma_{\alpha\rho}(z,y){\delta S\over \delta 
\sigma_{\alpha\beta} (z,x)}
F[\sigma ]>=0.
\label{42}
\end{equation}

Comparing this last equation with equation~(\ref{40}) 
for arbitrary $F[\sigma ]$ we obtain

\begin{equation}
\int d^{d}z \sigma_{\alpha\rho}(z,y)
{\delta lnJ\over\delta\sigma_{\alpha\beta} (z,x)}=
-(N+mL^{d}\delta^{d}(0))\delta_{\beta\rho}\delta^{d}(x-y),
\label{43}
\end{equation}

The solution to this equation is

\begin{equation}
lnJ=-(N+mL^{d}\delta^{d}(0))trTrln\sigma 
\label{44}
\end{equation}

The trace in the above equation runs over both Dirac 
and functional spaces. This is 
the main result of this article. 
As mentioned earlier the leading term of this Jacobian
has been obtained in~\cite{Ca}.
This 
Jacobian should be compared to the bosonic
one (\ref{14}). 

It is known that in models such as the Gross Neveu model 
an enlarged $O(2N)$ symmetry is present that is better exhibited
in terms of Majorana components~\cite{DHN}. One could have considered
a set of invariants 

\begin{equation}
\sigma'_{\alpha\beta} (x,y)=\psi_{\alpha}^{a}(x)\psi_{\beta}^{a}(y)
\label{45}
\end{equation}

\noindent for which the Jacobian is easily seen to be

\begin{equation}
lnJ=-{1\over 2}(N+mL^{d}\delta^{d}(0))trTrln\sigma
\label{46}
\end{equation}

Finaly if flavor degrees of freedom are added as in Nambu Jona-Lasinio
type models then one needs only include a further trace in flavor
space in the Jacobian (\ref{44}). 

\subsection{Effective action and large $N$ configuration.}

Our effective action is given by

\begin{equation}
S_{eff}=-ilnJ+NS
=iN trTrln\sigma + NS +imL^{d}\delta^{d}(0)Trln\sigma
=NS_{0}+S_{1}.
\label{47}
\end{equation}

We have assumed that we have again rescaled the fields
so that an overall factor of $N$ multiplies the action.
The leading $N\to\infty$ configuration is determined by

\begin{equation}
{\delta S_{0}\over\delta\sigma}|_{\sigma^{0}}=0
\label{48}
\end{equation}

Shifting about $\sigma^{0}$ as:

\begin{equation}
\sigma_{\alpha\beta}(x,y)=
\sigma_{\alpha\beta}^{0}(x,y)
+{1\over\sqrt{N}}\eta_{\alpha\beta}(x,y)
\label{79}
\end{equation}

\noindent we expand 

\begin{equation}
S_{eff}=NS_{0}(\sigma^{0})+S_{1}(\sigma^{0})-{i\over 2}C_{2}
+{1\over 2}D_{2}+\sum_{n=1}^{\infty}{1\over\sqrt{N}^{n}}\Big[
i(-1)^{n+1}[{C_{n+2}\over (n+2)} +mL^{d}\delta^{d}(0)
{C_{n}\over n}]+{1\over (n+2)!}D_{n+2}\Big]
\label{80}
\end{equation}

\noindent where

\begin{equation}
D_{n}=\int d^{d}x_{1} ... \int d^{d}x_{n}\int d^{d}y_{1}
... \int d^{d}y_{n} {\delta^{n}S\over
\delta\sigma_{\alpha_{1}\beta_{1}}(x_{1},y_{1}) ...
\delta\sigma_{\alpha_{n}\beta_{n}}(x_{n},y_{n})}|_{\sigma_{0}}
\eta_{\alpha_{1}\beta_{1}}(x_{1},y_{1}) ...
\eta_{\alpha_{n}\beta_{n}}(x_{n},y_{n})
\label{81}
\end{equation}

\noindent and

\begin{equation}
C_{n}=trTr((\sigma^{0})^{-1}\eta )^{n}
\label{82}
\end{equation}

At this point, a few comments are in order. The original fields
$\psi$ and $\bar{\psi}$ are fermionic fields, and
consequently they become Grassman variables under
path integral quantization. Using the Grassman nature of the
original variables, it follows that 
$\gamma^{0}_{\sigma\alpha}\sigma_{\sigma\beta}$ is 
an antihermitian bosonic
bilocal; all knowledge of the original fermionic statistics
is now coded in this property and 
the specific form of the interactions in
(\ref{80}). This provides a non trivial check of the 
effective field theory that can be 
carried out
with perturbation
theory.

\subsubsection{Perturbative Results}

We specialize now to the following Lagrangian density
(written before rescaling):

\begin{equation}
{\cal L} = \bar{\Psi}(i\not\!\partial-m)\Psi
+{\lambda^{2}\over 2N}(\bar{\Psi}\Psi)^{2}
+{\lambda_{5}^{2}\over 2N}(\bar{\Psi}\gamma_{5}\Psi)^{2}
\label{49}
\end{equation}

For now $\lambda$ and $\lambda_{5}$, 
assumed to be of the same order,
are left arbitrary, although 
later the cases $\lambda_{5}=0$ (the Gross Neveu model) and the
case $\lambda^{2} = -\lambda_{5}^{2}$ 
(a Nambu-Jona-Lasinio type model)
will be considered in detail.

By standard diagrammatic techniques, one can obtain for the two
point function

\begin{equation}
<\psi_{\alpha}(x)\bar{\psi}_{\beta}(y)>=\int
{d^{d}k\over (2\pi )^{d}} G^{(2)}_{\alpha\beta}(k)
e^{-ik(x-y)}
\label{83}
\end{equation}

\noindent with

\begin{equation}
\setlength{\unitlength}{1mm}
\begin{picture}(150,34)
\put(1,25){${G^{(2)}}_{\alpha\beta}(p)=N($}
\put(25,25){\line(1,0){12}}
\put(39,25){$+$}
\put(44.5,25){\line(1,0){12}}
\put(50.5,27){\circle{4}}
\put(50.5,25.5){\circle*{0.5}}
\put(50.5,24.5){\circle*{0.5}}
\put(62,25){$+$}
\put(70.5,25){\line(1,0){12}}
\put(73.5,27){\circle{4}}
\put(73.5,25.5){\circle*{0.5}}
\put(73.5,24.5){\circle*{0.5}}
\put(79.5,27){\circle{4}}
\put(79.5,25.5){\circle*{0.5}}
\put(79.5,24.5){\circle*{0.5}}
\put(90,26){$+$}
\put(97.5,25){\line(1,0){12}}
\put(103.5,27){\circle{4}}
\put(103.5,25.5){\circle*{0.5}}
\put(103.5,24.5){\circle*{0.5}}
\put(103.5,31){\circle{4}}
\put(103.5,29.5){\circle*{0.5}}
\put(103.5,28.5){\circle*{0.5}}
\put(118,25){$+O(\lambda^{6}))$}

\put(5,15){$+($}
\put(10.5,15){\line(1,0){12}}
\put(16.5,17){\circle{4}}
\put(14.5,15.5){\circle*{0.5}}
\put(18.5,15.5){\circle*{0.5}}
\put(29,15){$+$}
\put(36.5,15){\line(1,0){12}}
\put(39.5,17){\circle{4}}
\put(39.5,15.5){\circle*{0.5}}
\put(39.5,14.5){\circle*{0.5}}
\put(45.5,17){\circle{4}}
\put(47.5,15.5){\circle*{0.5}}
\put(43.5,15.5){\circle*{0.5}}
\put(53,15){$+$}
\put(60.5,15){\line(1,0){12}}
\put(63.5,17){\circle{4}}
\put(65.5,15.5){\circle*{0.5}}
\put(61.5,15.5){\circle*{0.5}}
\put(69.5,17){\circle{4}}
\put(69.5,15.5){\circle*{0.5}}
\put(69.5,14.5){\circle*{0.5}}

\put(78,16){$+$}
\put(84.5,15){\line(1,0){12}}
\put(90.5,17){\circle{4}}
\put(92.5,15.5){\circle*{0.5}}
\put(88.5,15.5){\circle*{0.5}}
\put(90.5,21){\circle{4}}
\put(90.5,19.5){\circle*{0.5}}
\put(90.5,18.5){\circle*{0.5}}

\put(102,16){$+$}
\put(108.5,15){\line(1,0){12}}
\put(114.5,17){\circle{4}}
\put(114.5,15.5){\circle*{0.5}}
\put(114.5,14.5){\circle*{0.5}}
\put(114.5,21){\circle{4}}
\put(112.5,19){\circle*{0.5}}
\put(116.5,19){\circle*{0.5}}

\put(8,5){$+$}
\put(15.5,6){\line(1,0){12}}
\put(21.5,6){\circle{4}}
\put(20,5.5){\circle*{0.5}}
\put(19,6.5){\circle*{0.5}}
\put(24,6.5){\circle*{0.5}}
\put(23,5.5){\circle*{0.5}}
\put(34,5){$+O(\lambda^{6})) +O({1\over N})$}
\end{picture}
\label{84}
\end{equation}

We have introduced a dot into our notation,
so as to indicate the contraction of Dirac and color indices.
Since the summation over color indices is taken into account 
in the overall factor of $N$, each dot should be thought of 
as a sum over the Dirac matrices ${\bf 1}$ and
$\gamma^{5}$, multiplied by the interaction strengths
$\lambda$ and $\lambda_{5}$ respectivly. For instance

\begin{equation}
\setlength{\unitlength}{1mm}
\begin{picture}(150,10)
\put(5,5){\line(1,0){12}}
\put(11,7){\circle{4}}
\put(11,5.5){\circle*{0.5}}
\put(11,4.5){\circle*{0.5}}
\put(17,5){$=
-i\sum_{M=\lambda{\bf 1},\lambda_{5}\gamma^{5}} 
\,\,{i\over \not\, p-m}
\int {d^{d}k\over (2\pi )^{d}}Tr\big[
M{i\over \not\, k-m}\big]M
{i\over \not\, p-m}$}
\end{picture}
\end{equation}

For completeness, every single diagram appearing in (\ref{84})
is explicitly written down in appendix A.

The connected piece of the four point function is given by:

\begin{eqnarray}
\nonumber
<\bar{\psi}^{a}_{\alpha'}(x_{1})&&
\psi^{a}_{\rho'}(y_{1})
\bar{\psi}^{b}_{\alpha}(x_{2})
\psi^{b}_{\rho}(y_{2})>
\quad =\quad\int {dp_{1}\over (2\pi )^{d}}\int {dp_{2}\over (2\pi )^{d}}
\int {dp_{3}\over (2\pi )^{d}}\int {dp_{4}\over (2\pi )^{d}}
(2\pi )^{d}\delta (p_{1}+p_{3}-p_{2}-p_{4})\times \\
&&\times e^{i(p_{3}x_{1}-p_{4}y_{1}+p_{1}x_{2}-p_{2}y_{2})}
{G^{(4)}}_{\alpha'\rho'\alpha\rho}^{conn}(p_{3},p_{4},p_{1},p_{2})
\label{85}
\end{eqnarray}

\noindent where

\begin{equation}
\setlength{\unitlength}{1.5mm}
\begin{picture}(110,10)
\put(5,5){${G^{(4)}}^{conn}(p_{1},p_{2},p_{3},p_{4})=N($}
\put(34,9){\line(1,-1){5}}
\put(34,4){\line(1,1){5}}
\put(36.5,6.85){\circle*{0.3}}
\put(36.5,6.15){\circle*{0.3}}
\put(43,5){$+$}
\put(46,7.75){\line(1,0){7.5}}
\put(46,5.25){\line(1,0){7.5}}
\put(49.5,8.1){\circle*{0.3}}
\put(49.5,7.4){\circle*{0.3}}
\put(49.5,5.6){\circle*{0.3}}
\put(49.5,4.9){\circle*{0.3}}
\put(49.5,6.5){\circle{2.5}}
\put(56,5){$+$}
\put(61,9){\line(1,-1){5}}
\put(61,4){\line(1,1){5}}
\put(63.5,6.85){\circle*{0.3}}
\put(63.5,6.15){\circle*{0.3}}
\put(62.75,7.7){\circle*{0.3}}
\put(62.45,7.2){\circle*{0.3}}
\put(63,8){\circle{1.5}}
\put(69,5){$+\quad permutations\quad +O(\lambda^{6}))+O(1).$}
\end{picture}
\label{86}
\end{equation}

Explicitly:

\begin{equation}
\setlength{\unitlength}{1.5mm}
\begin{picture}(110,10)
\put(5,9){\line(1,-1){5}}
\put(5,4){\line(1,1){5}}
\put(7.5,6.85){\circle*{0.3}}
\put(7.5,6.15){\circle*{0.3}}
\put(14,5){$=  i\sum_{M=\lambda{\bf 1},\lambda_{5}\gamma^{5}} 
\,\,[{i\over \not\, p_{4}-m}M{i\over \not\, p_{3}-m}]_{\rho'\alpha'}
[{i\over \not\, p_{2}-m}M{i\over \not\, p_{1}-m}]_{\rho\alpha}$}
\end{picture}
\nonumber
\end{equation}

\begin{equation}
\setlength{\unitlength}{1.5mm}
\begin{picture}(110,10)
\put(5,7.75){\line(1,0){7.5}}
\put(5,5.25){\line(1,0){7.5}}
\put(8.5,8.1){\circle*{0.3}}
\put(8.5,7.4){\circle*{0.3}}
\put(8.5,5.6){\circle*{0.3}}
\put(8.5,4.9){\circle*{0.3}}
\put(8.5,6.5){\circle{2.5}}
\put(14,5){$= -(i)^{2}\sum_{M,N=\lambda{\bf 1},\lambda_{5}\gamma^{5}} 
\,\,[{i\over \not\, p_{4}-m}M{i\over \not\, p_{3}-m}]_{\rho'\alpha'}
\int {d^{d}k\over (2\pi )^{d}}
Tr[{i\over \not\, k-m}
M{i\over \not\, k-\not\, p_{1}+\not\, p_{2}-m}N]
[{i\over \not\, p_{2}-m}N{i\over \not\, p_{1}-m}]_{\rho\alpha}$}
\end{picture}
\nonumber
\end{equation}

\begin{equation}
\setlength{\unitlength}{1.5mm}
\begin{picture}(110,10)
\put(5,9){\line(1,-1){5}}
\put(5,4){\line(1,1){5}}
\put(7.5,6.85){\circle*{0.3}}
\put(7.5,6.15){\circle*{0.3}}
\put(6.75,7.7){\circle*{0.3}}
\put(6.45,7.2){\circle*{0.3}}
\put(7,8){\circle{1.5}}
\put(14,5){$= -(i)^{2}\sum_{M,N=\lambda{\bf 1},\lambda_{5}\gamma^{5}} 
\,\,[{i\over \not\, p_{4}-m}
\int {d^{d}k\over (2\pi )^{d}}Tr[M{i\over \not\, k-m}]M
{i\over \not\, p_{4}-m}N{i\over \not\, p_{3}-m}]_{\rho'\alpha'}
[{i\over \not\, p_{2}-m}N{i\over \not\, p_{1}-m}]_{\rho\alpha}$}
\end{picture}
\label{three}
\end{equation}

\subsubsection{The Leading Configuration}

With a translationally invariant ansatz:

\begin{equation}
\sigma_{\alpha\beta} (x,y) = \int 
{d^{d}k\over (2\pi )^{d}} e^{ik(x-y)}\sigma_{\alpha\beta} (k)
\label{50}
\end{equation}

\noindent we obtain from (\ref{48})

\begin{equation}
-i{\sigma^{0}}^{-1}_{\rho\alpha}(k)=(\not\! k -m
+\lambda^{2}\tilde{\sigma}+\lambda_{5}^{2}\gamma_{5}
\tilde{\sigma_{5}})_{\alpha\rho}
\label{51}
\end{equation}

\noindent where

\begin{equation}
\tilde{\sigma}=\int {d^{d}k\over (2\pi)^{d}}\sigma_{\alpha\alpha}(k)
=-\tilde{G}=-\int {d^{d}k\over (2\pi )^{d}}Tr[G^{(2)}(k)]
\label{52}
\end{equation}

\noindent and

\begin{equation}
\tilde{\sigma_{5}}=\int {d^{d}k\over (2\pi)^{d}}
\gamma^{5}_{\alpha\beta}\sigma_{\alpha\beta}(k)
=-\tilde{G}_{5}=-\int {d^{d}k\over (2\pi )^{d}}
Tr[\gamma^{5}G^{(2)}(k)]
\label{53}
\end{equation}

This is the standard gap equation. We again assume that 
all integrals are regularized. Using (\ref{51}), we can
write, to $O(\lambda^{4})$

\begin{eqnarray}
\nonumber
G^{(2)}_{\alpha\beta}(p)&&=-\sigma^{0}_{\beta\alpha}(p)=
i(\not \! p -m-\lambda^{2}\tilde{G}-\lambda_{5}^{2}\gamma^{5}
\tilde{G}_{5})^{-1}_{\alpha\beta} \\
&&={i\over \not \! p-m}-i{i\over \not \! p-m}
(\lambda^{2}\tilde{G}+\lambda_{5}^{2}\gamma^{5}\tilde{G}_{5})
{i\over \not \! p-m}+(i)^{2}{i\over \not \! p-m}
(\lambda^{2}\tilde{G}+\lambda_{5}^{2}\gamma^{5}\tilde{G}_{5})
{i\over \not \! p-m}
(\lambda^{2}\tilde{G}+\lambda_{5}^{2}\gamma^{5}\tilde{G}_{5})
{i\over \not \! p-m}
\label{one}
\end{eqnarray}

Since

\begin{eqnarray}
\nonumber
\lambda^{2}\tilde{G}+\lambda_{5}^{2}\gamma^{5}\tilde{G}_{5}&&=
\sum_{M,N=\lambda{\bf 1},\lambda_{5}\gamma^{5}}
\int {d^{d}k\over (2\pi )^{d}}Tr[M{i\over \not \! k-m}]M-i
\sum_{M,N=\lambda{\bf 1},\lambda_{5}\gamma^{5}}
\int {d^{d}k\over (2\pi )^{d}}\int {d^{d}k'\over (2\pi )^{d}}
\times \\
&&\times
Tr[M{i\over \not \! k-m}N{i\over \not \! k-m}]
Tr[N{i\over \not \! k'-m}]M
+O(\lambda^{6}),
\end{eqnarray}

\noindent it is straightforward to substitute this expression into (\ref{one})
to obtain

\begin{equation}
\setlength{\unitlength}{1mm}
\begin{picture}(150,24)
\put(2,15){$G^{(2)}_{\alpha\beta}(p)=$}
\put(25,15){\line(1,0){12}}
\put(39,15){$+$}
\put(45,15){\line(1,0){12}}
\put(51,17){\circle{4}}
\put(51,15.5){\circle*{0.5}}
\put(51,14.5){\circle*{0.5}}
\put(62,15){$+$}
\put(71,15){\line(1,0){12}}
\put(74,17){\circle{4}}
\put(74,15.5){\circle*{0.5}}
\put(74,14.5){\circle*{0.5}}
\put(80,17){\circle{4}}
\put(80,15.5){\circle*{0.5}}
\put(80,14.5){\circle*{0.5}}
\put(90,16){$+$}
\put(98,15){\line(1,0){12}}
\put(104,17){\circle{4}}
\put(104,15.5){\circle*{0.5}}
\put(104,14.5){\circle*{0.5}}
\put(104,21){\circle{4}}
\put(104,19.5){\circle*{0.5}}
\put(104,18.5){\circle*{0.5}}
\put(118,15){$+O(\lambda^{6})$}
\end{picture}
\label{70}
\end{equation}

\noindent in agreement with the leading order term in the 
diagrammatic expansion for the two point
function.
Notice that in this case, not all diagrams contribute with the
same sign, due to the appearance of closed fermion loops in the 
leading order. It is remarkable that the original fermion statistics 
is completely accounted for by the change in sign of the leading order 
term in the Jacobian
by comparison to the bosonic case. 
Recall that 
$\gamma^{0}_{\alpha\tau}\sigma_{\alpha\beta}$ 
is an antihermitian bilocal
bosonic field. We will see that this change of sign correctly
reproduces the Fermi statistics to all orders of the $1/N$ expansion.

\subsubsection{Effective Field Theory Spectrum.}

In this section, we use the leading order $\sigma^{0}_{\alpha\beta}$,
at the level of quadratic fluctuations, to obtain the mass spectrum
of the effective field theory~\cite{JeRo2}. From (\ref{80}),
the leading quadratic
action is

\begin{eqnarray}
S_{2}&=&-{i\over 2}C_{2}+{1\over 2}D_{2} \\
\nonumber
&=& {1\over 2}\int dx_{1}dx_{2}dx_{3}dx_{4}\big[
(\lambda^{2}\delta_{\rho\alpha}
\delta_{\rho'\alpha'}
+\lambda_{5}^{2}\gamma^{5}_{\rho\alpha}
\gamma^{5}_{\rho'\alpha'})\delta (x_{1}-x_{2})
\delta (x_{1}-x_{3})\delta (x_{1}-x_{4})\\
\nonumber
&&-i[{\sigma^{0}}^{-1}(x_{1},x_{2})]_{\alpha\rho'}
[{\sigma^{0}}^{-1}(x_{3},x_{4})]_{\alpha'\rho}
\big]\eta_{\rho\alpha}(x_{4},x_{1})
\eta_{\rho'\alpha'}(x_{2},x_{3})
\label{87}
\end{eqnarray}

Now, using a symmetric Fourier transform

\begin{equation}
\eta_{\rho\alpha}(x_{1},x_{2})=\int {d^{d}p_{1}\over (2\pi )^{d/2}}
{d^{d}p_{2}\over (2\pi )^{d/2}}e^{i(p_{1}x_{1}-p_{2}x_{2})}
\eta_{\rho\alpha}(p_{1},p_{2})
\end{equation}

\noindent the quadratic action is

\begin{eqnarray}
S_{2}&=& {1\over 2}\int dp_{1}dp_{2}dp_{3}dp_{4}\big[
{\lambda^{2}\over (2\pi)^{d}}\delta_{\rho\alpha}
\delta_{\rho'\alpha'}\delta (p_{1}+p_{3}-p_{2}-p_{4})
+{\lambda_{5}^{2}\over (2\pi)^{d}}\gamma^{5}_{\rho\alpha}
\gamma^{5}_{\rho'\alpha'}\delta (p_{1}+p_{3}-p_{2}-p_{4})\\
\nonumber
&&-i\delta (p_{1}-p_{2})[{\sigma^{0}}^{-1}(p_{1})]_{\alpha\rho'}
\delta (p_{3}-p_{4})[{\sigma^{0}}^{-1}(p_{3})]_{\alpha'\rho}
\big]\eta_{\rho\alpha}(p_{4},p_{1})
\eta_{\rho'\alpha'}(p_{2},p_{3})
\end{eqnarray}

\noindent in momentum space. Using this action
we find that the propagator $A_{\alpha\rho\alpha'\rho'}$ of the field
$\sigma_{\alpha\rho}$ satisfies:

\begin{eqnarray}
\nonumber
\int dp_{1}dp_{2}\Big[
(\lambda^{2}\delta_{\mu\nu}\delta_{\rho\alpha}+
\lambda_{5}^{2}\gamma^{5}_{\mu\nu}\gamma^{5}_{\alpha\rho})
{\delta (k_{1}+p_{1}-k_{2}-p_{2})\over (2\pi )^{d}}
&&-i\delta (k_{1}-p_{2})
{\sigma^{0}_{\nu\alpha}}^{-1}(p_{1})
\delta (p_{1}-k_{2})
{\sigma^{0}_{\rho\mu}}^{-1}(p_{2}) \Big] 
\times \\
\times
A_{\alpha\rho\alpha'\rho'}(p_{1},p_{2},p_{3},p_{4}) 
&=&i\delta (k_{2}-p_{4})\delta (k_{1}-p_{3})
\delta_{\alpha'\mu}\delta_{\rho'\nu}
\label{88}
\end{eqnarray}

Using the argument preceding (\ref{56}), we make the ansatz

\begin{equation}
\setlength{\unitlength}{1mm}
\begin{picture}(150,20)
\put(0,15){$A_{\mu\nu\rho\tau}(p_{1},p_{2},p_{3},p_{4})
=\delta (p_{1}-p_{4})
\delta (p_{2}-p_{3})F_{\mu\nu\rho\tau}(p_{1},p_{2})+
\delta (p_{1}+p_{3}-p_{2}-p_{4})
G_{\mu\nu\rho\tau}(p_{1},p_{2},p_{3},p_{4})$}
\put(20,5){$=$}
\put(25,7){\line(1,0){4}}
\put(27,7){\circle*{1.5}}
\put(25,3){\line(1,0){4}}
\put(27,3){\circle*{1.5}}
\put(29,7){\line(1,-1){5}}
\put(29,3){\line(1,1){5}}
\put(38,5){$+$}
\put(44,7.5){\line(1,-1){5}}
\put(44,2.5){\line(1,1){5}}
\put(46.5,5){\circle*{3}}
\end{picture}
\label{89}
\end{equation}

Inserting this ansatz into (\ref{88}), we find the solution

\begin{eqnarray}
\nonumber
A_{\alpha'\rho'\alpha\rho}(p_{3},p_{4},p_{1},p_{2})&=&
-\delta (p_{3}-p_{2})\delta (p_{1}-p_{4})
\sigma_{\alpha'\rho}^{0}(p_{2})
\sigma_{\alpha\rho'}^{0}(p_{1})
+{i\over (2\pi )^{d}}\delta (p_{1}-p_{2}+p_{3}-p_{4})\times \\
&&\times\sum_{M,N=\lambda{\bf 1},\lambda_{5}\gamma^{5}}
[{\sigma^{0}}^{T}(p_{4})M{\sigma^{0}}^{T}(p_{3})]_{\rho'\alpha'}
\Gamma_{MN}(p_{2}-p_{1})
[{\sigma^{0}}^{T}(p_{2})N{\sigma^{0}}^{T}(p_{1})]_{\rho\alpha} 
\label{90}
\end{eqnarray}

\noindent where

\begin{equation}
\nonumber
\Gamma_{MN}(p_{2}-p_{1})=[{\bf 1}+iG(p_{2}-p_{1})]^{-1}_{MN}
\end{equation}

\noindent and

\begin{equation}
G_{MN}(p_{2}-p_{1})=\int {d^{d}k\over (2\pi )^{d}}
Tr[M{\sigma^{0}}^{T}(k+p_{2}-p_{1})
N{\sigma^{0}}^{T}(k)]=G_{NM}(p_{1}-p_{2})
\label{four}
\end{equation}

This last term corresponds to summing up the bubble diagrams,
which we know dominate the large $N$ limit. 
The diagonal elements of $\Gamma_{MN}$
are identical to the propagators of the auxiliary fields which
are usually employed to study this model~\cite{Gross}. 
The first term, consists of disconnected diagrams in terms 
of the original fermion fields and has 
no natural counterpart in the auxiliary field approach. 
This term is however, crucial to obtain a systematic expansion. This
will already be evident at the perturbative level, in the discussion
of the next section. 
It is easy to see how the Fermi statistics is reflected in the above
propagator:
The first term corresponds to an exchange of
the two ingoing fermions, with respect to the leading configuration.
It thus appears with the opposite sign to the corresponding term in 
the bosonic propagator (\ref{56}). Also, the fermionic
bubble in the denominator of the second term appears with a minus
sign, reflecting the fact that closed fermion loops come with a
factor of $-1$.

A full nonperturbative treatment of the spectrum will be discussed
in a later section. At this point we simply construct the connected
piece of $A_{\alpha'\rho'\alpha\rho}$ to $O({\lambda^{4}})$. 
First, expand the second (connected) term of (\ref{90}) as

\begin{eqnarray}
A_{\alpha'\rho'\alpha\rho}^{conn}(&&p_{3},p_{4},p_{1},p_{2})=
{i\over (2\pi )^{d}}\delta (p_{1}-p_{2}+p_{3}-p_{4}) 
\sum_{M=\lambda{\bf 1},\lambda_{5}\gamma^{5}}
[{\sigma^{0}}^{T}(p_{4})M{\sigma^{0}}^{T}(p_{3})]_{\rho'\alpha'}
[{\sigma^{0}}^{T}(p_{2})M{\sigma^{0}}^{T}(p_{1})]_{\rho\alpha} \\
\nonumber
&-&{(i)^{2}\over (2\pi )^{d}}\delta (p_{1}-p_{2}+p_{3}-p_{4}) 
\sum_{M,N=\lambda{\bf 1},\lambda_{5}\gamma^{5}}
[{\sigma^{0}}^{T}(p_{4})M{\sigma^{0}}^{T}(p_{3})]_{\rho'\alpha'}
G_{MN}(p_{2}-p_{1})
[{\sigma^{0}}^{T}(p_{2})N{\sigma^{0}}^{T}(p_{1})]_{\rho\alpha} 
+O(\lambda^{6}).
\label{two}
\end{eqnarray}

Using the expansion for 
$({\sigma^{0}})^{T}_{\beta\alpha}=\sigma^{0}_{\alpha\beta}
=-G^{(2)}_{\alpha\beta}$ to $O(\lambda^{2})$ 
obtained in a previous section, one readily verifies that the
above expression is in exact agreement with the expansion 
(\ref{86}-\ref{three}).

\subsubsection{Corrections to the one and two point functions.}

In this section we compute the tadpole, (subleading) quadratic, cubic and 
quartic interaction vertices of the effective field theories.
We use these vertices in this section to obtain agreement 
with the perturbative results quoted in subsection $2$ for
the (fermion) two point function. 
These vertices will be used in a calculation
in a later section.

From the following terms in the action

\begin{eqnarray}
\nonumber
S_{tp} &=& imL^{d}\delta^{d}(0){1\over\sqrt{N}}C_{1} \\
\nonumber
S_{c} &=& i{1\over 3\sqrt{N}}C_{3} \\
\nonumber
S_{sq} &=& -imL^{d}\delta^{d}(0){1\over 2N}C_{2} \\
S_{q} &=& -i{1\over 4N}C_{4}
\label{93}
\end{eqnarray}

\noindent we easily derive the rules

\begin{equation}
\setlength{\unitlength}{1mm}
\begin{picture}(150,68)
\put(10,63){\line(0,1){5}}
\put(10,67){\circle*{2}}
\put(12,63){$p_{1},p_{2}$}
\put(12,66){$\alpha ,\beta$}
\put(35,64){$-{mL^{d}\delta^{d}(0)\over\sqrt{N}}
[{\sigma^{0}}^{-1}]_{p_{2}\beta p_{1}\alpha}$}

\put(5,51){\line(1,0){10}}
\put(10,51){\circle*{2}}
\put(1,55){$\alpha \beta$}
\put(0,52){$k_{1},k_{2}$}
\put(14,55){$\gamma \nu$}
\put(13,52){$p_{1},p_{2}$}
\put(35,52){${mL^{d}\delta^{d}(0)\over 2N}
[{\sigma^{0}}^{-1}]_{k_{2}\beta p_{1}\gamma}
[{\sigma^{0}}^{-1}]_{p_{2}\nu k_{1}\alpha}$}

\put(5,36){\line(1,0){10}}
\put(10,36){\line(0,1){5}}
\put(1,36){$\gamma \nu$}
\put(0,33){$k_{1},k_{2}$}
\put(14,36){$\phi \psi$}
\put(13,33){$p_{1},p_{2}$}
\put(7,43){$\alpha \beta$}
\put(6,41){$q_{1},q_{2}$}
\put(35,40){${-1\over 3!\sqrt{N}}(
[{\sigma^{0}}^{-1}]_{k_{2}\nu p_{1}\phi}
[{\sigma^{0}}^{-1}]_{p_{2}\psi q_{1}\alpha}
[{\sigma^{0}}^{-1}]_{q_{2}\beta k_{1}\gamma}
+[{\sigma^{0}}^{-1}]_{p_{2}\psi k_{1}\gamma}
[{\sigma^{0}}^{-1}]_{k_{2}\nu q_{1}\alpha}
[{\sigma^{0}}^{-1}]_{q_{2}\beta p_{1}\phi})$}

\put(6,14){\line(1,1){8}}
\put(14,14){\line(-1,1){8}}
\put(1,13.5){$\phi \psi$}
\put(0,10.5){$k_{1},k_{2}$}
\put(14,13.5){$\mu \tau$}
\put(13,10.5){$p_{1},p_{2}$}
\put(1,26.5){$\alpha \beta$}
\put(0,23.5){$q_{1},q_{2}$}
\put(14,26.5){$\gamma \nu$}
\put(13,23.5){$l_{1},l_{2}$}

\put(35,29){${1\over 4!N}(
[{\sigma^{0}}^{-1}]_{k_{2}\psi p_{1}\mu}
[{\sigma^{0}}^{-1}]_{p_{2}\tau q_{1}\alpha}
[{\sigma^{0}}^{-1}]_{q_{2}\beta l_{1}\gamma}
[{\sigma^{0}}^{-1}]_{l_{2}\nu k_{1}\phi}
+[{\sigma^{0}}^{-1}]_{k_{2}\psi p_{1}\mu}
[{\sigma^{0}}^{-1}]_{p_{2}\tau l_{1}\gamma}
\times$}

\put(35,21){$\times
[{\sigma^{0}}^{-1}]_{l_{2}\nu q_{1}\alpha}
[{\sigma^{0}}^{-1}]_{q_{2}\beta k_{1}\phi}+
[{\sigma^{0}}^{-1}]_{k_{2}\psi q_{1}\alpha}
[{\sigma^{0}}^{-1}]_{q_{2}\beta l_{1}\gamma}
[{\sigma^{0}}^{-1}]_{l_{2}\nu p_{1}\mu}
[{\sigma^{0}}^{-1}]_{p_{2}\tau k_{1}\phi}+$}

\put(35,13){$+
[{\sigma^{0}}^{-1}]_{k_{2}\psi q_{1}\alpha}
[{\sigma^{0}}^{-1}]_{q_{2}\beta p_{1}\mu}
[{\sigma^{0}}^{-1}]_{p_{2}\tau l_{1}\gamma}
[{\sigma^{0}}^{-1}]_{l_{2}\nu k_{1}\phi}
+[{\sigma^{0}}^{-1}]_{k_{2}\psi l_{1}\gamma}
[{\sigma^{0}}^{-1}]_{l_{2}\nu q_{1}\alpha}
\times$}

\put(35,5){$\times
[{\sigma^{0}}^{-1}]_{q_{2}\beta p_{1}\mu}
[{\sigma^{0}}^{-1}]_{p_{2}\tau k_{1}\phi}
+[{\sigma^{0}}^{-1}]_{k_{2}\psi l_{1}\gamma}
[{\sigma^{0}}^{-1}]_{l_{2}\nu p_{1}\mu}
[{\sigma^{0}}^{-1}]_{p_{2}\tau q_{1}\alpha}
[{\sigma^{0}}^{-1}]_{q_{2}\beta k_{1}\phi})$}

\end{picture}
\label{94}
\end{equation}

We have again gone over to an obvious matrix notation. 
Since $\sigma^{0}$ is diagonal in momentum space, we have

\begin{equation}
\nonumber
[{\sigma^{0}}^{-1}]_{p\alpha k\beta}=
\delta (p-k) [{\sigma^{0}}^{-1} (p)]_{\alpha\beta}
\end{equation}

As in the 
bosonic case, two processes contribute to the $O({1\over\sqrt{N}})$
correction to the one point function: the cubic tadpole and the linear
tadpole. Using the expression (\ref{90}) for the propagator, and 
using the expansion $(93)$ (for the second connected term
of the propagator), together with the above rules, we write down
expressions for the cubic tadpole

\begin{equation}
\setlength{\unitlength}{1mm}
\begin{picture}(150,45)
\put(19,32){\line(0,1){8}}
\put(17.75,31.25){$\times$}
\put(19,42){\circle{4}}
\put(24,36){$=-{1\over\sqrt{N}}\int dp_{1}dp_{2}dp_{3}
 {\sigma^{0}}^{-1}_{\epsilon'\gamma}(p_{1})
 {\sigma^{0}}^{-1}_{\gamma'\epsilon}(p_{2})
 {\sigma^{0}}^{-1}_{\delta\delta'}(p_{3})
 A_{\gamma\gamma'\epsilon\delta}(p_{1},p_{2},p_{2},p_{3})
 A_{\delta'\epsilon'\alpha\beta}(p_{3},p_{1},k_{1},k_{2})$}
\put(24,26){$={mL^{d}\delta^{d}(0)\over\sqrt{N}}
\int d^{d}p {\sigma^{0}}^{-1}_{\rho\epsilon}(p)
A_{\epsilon\rho\alpha\beta}(p,p,k_{1},k_{2})
+{1\over\sqrt{N}}\delta (k_{1}-k_{2})\Big[
i\int{dp\over (2\pi )^{d}}
\sum_{M=\lambda{\bf 1},\lambda_{5}\gamma^{5}}$}
\put(24,16){$[{\sigma^{0}}^{T}(k_{2})M{\sigma^{0}}^{T}(p)
M{\sigma^{0}}^{T}(k_{1})]_{\beta\alpha}
+ \sum_{M,N=\lambda{\bf 1},\lambda_{5}\gamma^{5}}
\int{dp\over (2\pi )^{d}}[{\sigma^{0}}^{T}
(k_{2})N{\sigma^{0}}^{T}(p)
M{\sigma^{0}}^{T}(k_{1})]_{\beta\alpha}\times$}
\put(25,6){$G_{MN}(k_{2}-p)
+\int{dp_{1}\over (2\pi )^{d}}
\int{dp_{2}\over (2\pi )^{d}}
Tr[M{\sigma^{0}}^{T}(p_{1})N{\sigma^{0}}^{T}(p_{1})
M{\sigma^{0}}^{T}(p_{2})] [{\sigma^{0}}^{T}
(k_{2})N{\sigma^{0}}^{T}(k_{1})]_{\beta\alpha}$}
\end{picture}
\label{95}
\end{equation}

\noindent to $O(\lambda^{6})$ and for the linear tadpole

\begin{equation}
\setlength{\unitlength}{1mm}
\begin{picture}(150,15)
\put(19,2){\line(0,1){8}}
\put(17.75,1.25){$\times$}
\put(19,10){\circle*{2}}
\put(24,6){$=-{mL^{d}\delta^{d}(0)\over\sqrt{N}}
\int d^{d}p {\sigma^{0}}^{-1}_{\rho\epsilon}(p)
A_{\epsilon\rho\alpha\beta}(p,p,k_{1},k_{2})$}
\end{picture}
\label{96}
\end{equation}

Again we find that the linear tadpole is essential to cancel momentum
dependent infinities arising from the integration in the cubic loop.
The sum of the two tadpoles is given by

\begin{eqnarray}
\nonumber
<\eta_{\alpha\beta}&&(k_{1},k_{2}) >
={1\over\sqrt{N}}\delta (k_{1}-k_{2})\Big[
i\int{dp\over (2\pi )^{d}}
\sum_{M=\lambda{\bf 1},\lambda_{5}\gamma^{5}}
[{\sigma^{0}}^{T}(k_{2})M{\sigma^{0}}^{T}(p)
M{\sigma^{0}}^{T}(k_{1})]_{\beta\alpha}
+ \sum_{M,N=\lambda{\bf 1},\lambda_{5}\gamma^{5}}
\int{dp\over (2\pi )^{d}}\times \\
\nonumber
&&\times [{\sigma^{0}}^{T}(k_{2})N{\sigma^{0}}^{T}(p)
M{\sigma^{0}}^{T}(k_{1})]_{\beta\alpha}G_{MN}(k_{2}-p)
+\int{dp_{1}\over (2\pi )^{d}}
\int{dp_{2}\over (2\pi )^{d}}
Tr[N{\sigma^{0}}^{T}(p_{1})M{\sigma^{0}}^{T}(p_{1})
N{\sigma^{0}}^{T}(p_{2})] \times \\
&&\times [{\sigma^{0}}^{T}(k_{2})
M{\sigma^{0}}^{T}(k_{1})]_{\beta\alpha}+O(\lambda^{6})
\end{eqnarray}

Substituting the expressions for $\sigma^{0}$ 
and $G_{MN}$ given in
(\ref{70}) and (\ref{four}), 
we find complete agreement with the $O(1)$ term of
(\ref{84}).
Notice that once again all diagrams enter with the correct signs - thus
corrections computed using our Jacobian (\ref{44}) have Fermi statistics 
correctly accounted for! 
(It should be recalled that 
$<\sigma_{\alpha\beta}(k_{1},k_{2})>=-G^{(2)}_{\beta\alpha}(k_{1},k_{2})$).

\section{A non perturbative analysis of the effective field theory
spectrum}

In this section, we obtain the non perturbative
effective field theory spectrum and propagator
for the Gross-Neveu
model and a Nambu-Jona-Lasinio type model
(a model with a continuos chiral symmetry)
and relate them to previously obtained results.

\subsection {The Gross-Neveu Model}
For the Gross-Neveu model, we set $\lambda_{5}=m=0$
and work in $d=1+1$ dimensions

\begin{equation}
{\cal L}=\bar{\psi}^{a}i\not\!\partial\psi^{a}
+{\lambda^{2}\over 2N}(\bar{\psi}^{a}\psi^{a})^{2}
\label{400}
\end{equation}

Notice that this model has a discrete chiral symmetry

\begin{equation}
\psi^{a}_{\alpha}\to\gamma^{5}_{\alpha\nu}\psi^{a}_{\nu}\quad
\bar{\psi}^{a}_{\alpha}\to
-\bar{\psi}^{a}_{\nu}\gamma^{5}_{\nu\alpha}\quad
\label{401}
\end{equation}

Using the formalism developed in section $III$, we find that the
leading configuration $\sigma^{0}$ satisfies

\begin{equation}
-i[{\sigma^{0}}^{-1}]_{\rho\alpha}(k)=
(\not\! k+\lambda^{2}\sigma)_{\alpha\rho}
\label{402}
\end{equation}

\noindent where

\begin{equation}
\sigma=\int{d^{2}k\over (2\pi )^{2}}Tr[\sigma (k)]
\label{403}
\end{equation}

Inserting (\ref{402}) into (\ref{403}) we find the standard result

\begin{equation}
\sigma =\int{d^{2}k\over (2\pi )^{2}}
Tr[\not\! k +\lambda^{2}\sigma]^{-1} =
{\lambda^{2}\sigma\over 2\pi}log {\Lambda^{2}\over
(\lambda^{2}\sigma)^{2}}
\label{404}
\end{equation}

\noindent where $\Lambda$ is an ultraviolet cut off. Therefore

\begin{equation}
\lambda^{2}\sigma=\Lambda e^{-{\pi\over\lambda^{2}}}
\label{405}
\end{equation}

Thus the fermions have aquired a dynamically generated mass
$\lambda^{2}\sigma$ and the discrete chiral symmetry
(\ref{401}) has been spontaneously broken, 
as is well known~\cite{Gross}.
This mass can be made finite as
$\Lambda\to\infty$ by adding the
counter term

\begin{equation}
S_{ct}=-{c\over 2}\int {d^{2}p_{1}\over 2\pi }
{d^{2}p_{2}\over 2\pi }d^{2}p_{3} d^{2}p_{4}
\delta (p_{1}+p_{3}-p_{2}-p_{4}) 
\eta_{\alpha\alpha}(p_{1},p_{2})
\eta_{\nu\nu}(p_{3},p_{4})
\label{407}
\end{equation}

\noindent to our effective action, corresponding to a coupling
constant renormalization of the original theory. Fixing the
renormalized mass to some value $m_{r}$

\begin{equation}
m_{r}=\Lambda e^{-{\pi\over \lambda^{2} - c}}
=\Lambda e^{-{\pi\over \lambda_{0}^{2}}}
\label{408}
\end{equation}

\noindent we can fix $c$

\begin{equation}
c=\lambda^{2} +[{1\over 2\pi}log {m_{r}^{2}\over \Lambda^{2}}]^{-1}
\label{409}
\end{equation}

Using the arguments constructed in section B3 we know that the
effective field theory propagator

\begin{equation}
A_{\mu\nu\mu'\nu'}(p_{1},p_{2},p_{3},p_{4})
=<\eta_{\mu\nu}(p_{1},p_{2})
\eta_{\mu'\nu'}(p_{3},p_{4})>
\label{411}
\end{equation}

\noindent can be expressed in
terms of the leading configuration $\sigma^{0}$
as

\begin{eqnarray}
\nonumber
A_{\mu\nu\mu'\nu'}&&(p_{1},p_{2},p_{3}p_{4})=
-\delta(p_{3}-p_{2})\delta(p_{1}-p_{4})
\sigma^{0}_{\mu\nu'}(p_{1})\sigma^{0}_{\mu'\nu}(p_{3})\\
&&+{i\lambda_{0}^{2}\over (2\pi )^{2}}\delta (p_{1}+p_{3}-p_{2}-p_{4})
{\sigma^{0}_{\mu\rho}(p_{1})\sigma^{0}_{\rho\nu}(p_{2})
\sigma^{0}_{\mu'\tau}(p_{3})\sigma^{0}_{\tau\nu'}(p_{4})\over
1+i\lambda_{0}^{2}\int {d^{2}k\over (2\pi )^{2}}
\sigma^{0}_{\sigma\gamma}(k)
\sigma^{0}_{\gamma\sigma}(k-p_{1}+p_{2})}
\label{412}
\end{eqnarray}

To construct
the propagator explicitly, we perform the integral~\cite{Gross}:

\begin{eqnarray}
\nonumber
I &&= i\lambda_{0}^{2}\int {d^{2}l\over (2\pi )^{2}}
\sigma^{0}_{\sigma\gamma}(l)
\sigma^{0}_{\gamma\sigma}(l-p_{1}+p_{2})\\
&&= -{\lambda_{0}^{2}\over 2\pi }
\sqrt{4m_{r}^{2}-k^{2}\over -k^{2}}log\big[
{\sqrt{4m_{r}^{2}-k^{2}}-\sqrt{-k^{2}}\over
\sqrt{4m_{r}^{2}-k^{2}} +\sqrt{-k^{2}}}\big]
-{\lambda_{0}^{2}\over \pi}log {\Lambda\over m_{r}}.
\label{413}
\end{eqnarray}

\noindent where $k=p_{1}-p_{2}$.
Now, from (\ref{408}) we see that

\begin{equation}
{\lambda_{0}^{2}\over\pi} log{\Lambda\over m_{r}}=1
\label{414}
\end{equation}

so that

\begin{equation}
I = -{\lambda_{0}^{2}\over 2\pi }
\sqrt{4m_{r}^{2}-k^{2}\over -k^{2}}log\big[
{\sqrt{4m_{r}^{2}-k^{2}}-\sqrt{-k^{2}}\over
\sqrt{4m_{r}^{2}-k^{2}} +\sqrt{-k^{2}}}\big]
-1.
\label{415}
\end{equation}

Inserting this into the propagator (\ref{412}), we obtain

\begin{eqnarray}
\nonumber
A_{\mu\nu\mu'\nu'}&&(p_{1},p_{2},p_{3}p_{4})=
-\delta (p_{2}-p_{3})\delta (p_{4}-p_{1})
[{i\over \not\! p_{1}-m_{r}}]_{\nu'\mu}
[{i\over \not\! p_{3}-m_{r}}]_{\nu\mu'}\\
&&-{i\over 2\pi}
[{i\over \not\! p_{2}-m_{r}}{i\over \not\! p_{1}-m_{r}}]_{\nu\mu}
[{i\over \not\! p_{4}-m_{r}}{i\over \not\! p_{3}-m_{r}}]_{\nu'\mu'}
{\delta (p_{1}+p_{3}-p_{2}-p_{4})\over
\sqrt{4m_{r}^{2}-k^{2}\over -k^{2}}log\big[
{\sqrt{4m_{r}^{2}-k^{2}}-\sqrt{-k^{2}}\over
\sqrt{4m_{r}^{2}-k^{2}}+\sqrt{-k^{2}}}\big]}
\label{416}
\end{eqnarray}

\noindent Notice that all 
dependence on $\Lambda$ has disappeared as it must.

The interpretation of the above propagator is clearer in terms
of rapidity variables
$p^{0}=m cosh(\theta),p^{1}=m sinh(\theta)$. If $\phi=\theta_{1}-\theta_{2}$
the connected piece of the propagator 
(after the removal of the external legs) is given by

\begin{equation}
D(\phi )=2\pi i {tanh({\phi\over 2})\over\phi}.
\label{vv}
\end{equation}

For on-shell particle-antiparticle scattering (corresponding to time flowing
from right to left in (89)) we let $\phi\to\phi-i\pi$
($p_{1}\to p_{1},p_{2}\to -p_{2}$) and obtain the amplitude 

\begin{equation}
D(\phi )=2\pi i {coth({\phi\over 2})\over \phi-i\pi}.
\label{pvv}
\end{equation}

Since $s=(p_{1}+p_{2})^{2}=4m^{2}cosh^{2}({\phi\over 2})$
we see that as $\phi\to 0$ the above amplitude displays a cut at
$s=4m^{2}$.

\subsection {A Nambu-Jona-Lasinio type model}

The model we consider in this section has 
$\lambda_{5}^{2}=-\lambda^{2}$
and $m=0$

\begin{equation}
{\cal L}=\bar{\psi}^{a}i\not\!\partial\psi^{a}
+{\lambda^{2}\over 2N}(\bar{\psi}^{a}\psi^{a})^{2}
-{\lambda_{5}^{2}\over 2N}(\bar{\psi}^{a}\gamma^{5}\psi^{a})^{2}
\label{418}
\end{equation}

We again work in $d=1+1$ dimensions.
This model is invariant under the global continuos
chiral transformation

\begin{equation}
\psi^{a}_{\alpha}\to
{e^{\alpha\gamma^{5}}}_{\alpha\nu}\psi^{a}_{\nu}\quad
\bar{\psi}^{a}_{\alpha}\to
-\bar{\psi}^{a}_{\nu}
{e^{\alpha\gamma^{5}}}_{\nu\alpha}\quad
\label{419}
\end{equation}

The leading configuration $\sigma^{0}$ satisfies

\begin{equation}
-i[{\sigma^{0}}^{-1}]_{\rho\alpha}(k)=
(\not\! k+\lambda^{2}\sigma
+\lambda_{5}^{2}\gamma^{5}\tilde{\sigma})_{\alpha\rho}
\label{420}
\end{equation}

\noindent where

\begin{equation}
\sigma=\int{d^{2}k\over (2\pi )^{2}}Tr[\sigma (k)]
\label{421}
\end{equation}

\noindent and

\begin{equation}
\tilde{\sigma}=i\int{d^{2}k\over (2\pi )^{2}}
Tr[\gamma^{5}\tilde{\sigma}(k)]
\label{422}
\end{equation}

Inserting (\ref{420}) into (\ref{421}) yields the
standard gap equation~\cite{NaJo}

\begin{equation}
\sigma=\int{d^{2}k\over (2\pi )^{2}}
[\not\! k+\lambda^{2}\gamma^{5}\tilde{\sigma}
+\lambda^{2}\tilde{\sigma}]^{-1}_{\nu\nu}=
\int{d^{2}k\over (2\pi )^{2}}
[{4i\lambda^{2}\sigma\over k^{2}
-4\lambda^{4}(\sigma^{2}+\tilde{\sigma}^{2})}]
\label{423}
\end{equation}

Similarily, the gap equation for
$\tilde{\sigma}$~\cite{NaJo} is
obtained by inserting (\ref{420}) into (\ref{422})

\begin{equation}
\tilde{\sigma}=\int{d^{2}k\over (2\pi )^{2}}
[{-i\gamma^{5}\over\not\! k
+\lambda^{2}\gamma^{5}\tilde{\sigma}
+\lambda^{2}\tilde{\sigma}}]_{\nu\nu}=
\int{d^{2}k\over (2\pi )^{2}}
[{4i\lambda^{2}\tilde{\sigma}\over k^{2}
-4\lambda^{4}(\sigma^{2}+\tilde{\sigma}^{2})}]
\label{424}
\end{equation}

Fixing the mass to some value $m_{r}$, and parametrizing~\cite{NaJo}

\begin{equation}
2\lambda^{2}\sigma =m_{r}cos\theta\quad
2\lambda^{2}\tilde{\sigma}=m_{r}sin\theta,
\label{426}
\end{equation}

\noindent we find

\begin{equation}
{1\over 2\lambda^{2}}={1\over 2\pi}log{\Lambda^{2}\over m_{r}^{2}}
\label{427}
\end{equation}

The expression for the
effective field theory propagator in terms of the leading
configuration is

\begin{eqnarray}
\nonumber
A_{\mu\nu\mu'\nu'}&&(p_{1},p_{2},p_{3}p_{4})=
-\delta(p_{3}-p_{2})\delta(p_{1}-p_{4})
\sigma^{0}_{\mu\nu'}(p_{1})\sigma^{0}_{\mu'\nu}(p_{3})\\
&&+{i\lambda_{0}^{2}\over (2\pi )^{2}}\delta (p_{1}+p_{3}-p_{2}-p_{4})
\sum_{M,N={\bf 1},\gamma^{5}}
{[{\sigma^{0}}^{T}(p_{1})M{\sigma^{0}}^{T}(p_{2})]_{\nu\mu}
[{\sigma^{0}}^{T}(p_{3})N{\sigma^{0}}^{T}(p_{4})]_{\nu'\mu'}\over
1+i\lambda_{0}^{2}\int {d^{2}k\over (2\pi )^{2}}
Tr[{\sigma^{0}}^{T}(k)M
{\sigma^{0}}^{T}(k-p_{1}+p_{2})N]}
\label{428}
\end{eqnarray}

To make the following arguments simple and transparent, we
pick $\theta =0$ in $\sigma^{0}_{\alpha\beta}$. To obtain
an explicit expression for the propagator, we need to compute
three different integrals. The first integral was considered
in the previous section, so we simply quote the result~\cite{Gross}

\begin{eqnarray}
\nonumber
I_{1} &&= i\lambda_{0}^{2}\int {d^{2}k\over (2\pi )^{2}}
\sigma^{0}_{\sigma\gamma}(k)
\sigma^{0}_{\gamma\sigma}(k-p_{1}+p_{2})\\
&&= -{\lambda_{0}^{2}\over 2\pi }
\sqrt{4m_{r}^{2}-k^{2}\over -k^{2}}log\big[
{\sqrt{4m_{r}^{2}-k^{2}}-\sqrt{-k^{2}}\over
\sqrt{4m_{r}^{2}-k^{2}} +\sqrt{-k^{2}}}\big]-1
\label{429}
\end{eqnarray}

The second integral

\begin{equation}
I_{2}=\int {d^{2}k\over (2\pi )^{2}}Tr
[{\sigma^{0}}^{T}(k-p_{1}+p_{2}){\bf 1}
{\sigma^{0}}^{T}(k)\gamma^{5}]
\label{430}
\end{equation}

\noindent vanishes, due to the trace over
Dirac indices. The third integral
is easily computed

\begin{eqnarray}
\nonumber
I_{3}&&=i\lambda_{0}^{2}\int {d^{2}k\over (2\pi )^{2}}Tr
[{\sigma^{0}}^{T}(k-p_{1}+p_{2})i\gamma^{5}
{\sigma^{0}}^{T}(k)i\gamma^{5}]\\
&&= -{\lambda_{0}^{2}\over 2\pi }
\sqrt{4m_{r}^{2}-k^{2}\over -k^{2}}log\big[
{\sqrt{4m_{r}^{2}-k^{2}}-\sqrt{-k^{2}}\over
\sqrt{4m_{r}^{2}-k^{2}} +\sqrt{-k^{2}}}\big]
-{\lambda_{0}^{2}\over 2\pi }log{\Lambda^{2}\over m_{r}^{2}}
+{\lambda_{0}^{2}\over\pi}
\label{431}
\end{eqnarray}

where $\Lambda$ is an ultraviolet cut off. Using the renormalization
condition, in the form (\ref{427}), we may rewrite this as

\begin{equation}
I_{3}=-{\lambda_{0}^{2}\over 2\pi }
\sqrt{4m_{r}^{2}-k^{2}\over -k^{2}}log\big[
{\sqrt{4m_{r}^{2}-k^{2}}-\sqrt{-k^{2}}\over
\sqrt{4m_{r}^{2}-k^{2}} +\sqrt{-k^{2}}}\big]
-1+{\lambda_{0}^{2}\over\pi}
\label{432}
\end{equation}

Thus, the effective field theory propagator reads explicitly

\begin{eqnarray}
\nonumber
A_{\mu\nu\mu'\nu'}&&(p_{1},p_{2},p_{3}p_{4})=
-\delta (p_{2}-p_{3})\delta (p_{4}-p_{1})
[{i\over \not\! p_{1}-m_{r}}]_{\nu'\mu}
[{i\over \not\! p_{3}-m_{r}}]_{\nu\mu'}\\
\nonumber
&&-{i\over 2\pi}
[{i\over \not\! p_{2}-m_{r}}{i\over \not\! p_{1}-m_{r}}]_{\nu\mu}
[{i\over \not\! p_{4}-m_{r}}{i\over \not\! p_{3}-m_{r}}]_{\nu'\mu'}
{\delta (p_{1}+p_{3}-p_{2}-p_{4})\over
\sqrt{4m_{r}^{2}-k^{2}\over -k^{2}}log\big[
{\sqrt{4m_{r}^{2}-k^{2}}-\sqrt{-k^{2}}\over
\sqrt{4m_{r}^{2}-k^{2}}+\sqrt{-k^{2}}}\big]}\\
&&-{i\over 2\pi}
[{i\over \not\! p_{2}-m_{r}}{i\over \not\! p_{1}-m_{r}}]_{\nu\mu}
[{i\over \not\! p_{4}-m_{r}}{i\over \not\! p_{3}-m_{r}}]_{\nu'\mu'}
{\delta (p_{1}+p_{3}-p_{2}-p_{4})\over
\sqrt{4m_{r}^{2}-k^{2}\over -k^{2}}log\big[
{\sqrt{4m_{r}^{2}-k^{2}}-\sqrt{-k^{2}}\over
\sqrt{4m_{r}^{2-k^{2}}}+\sqrt{-k^{2}}}\big]-2}
\label{433}
\end{eqnarray}

The first term again represents the crossed
propagation of two free fermions.
The second term has been discussed earlier.
The third term
above, is however new. It has the property that

\begin{equation}
\sqrt{4m_{r}^{2}-k^{2}\over -k^{2}}log\big[
{\sqrt{4m_{r}^{2}-k^{2}}-\sqrt{-k^{2}}\over
\sqrt{4m_{r}^{2}-k^{2}}
+\sqrt{-k^{2}}}\big]|_{k^{2}=0}=2.
\label{434}
\end{equation}

This has been interpreted as the signature
of a massless scalar in the spectrum of the
model~\cite{Sweica}. This massless particle is  
associated with the fact that
the continuos chiral symmetry of the model, is
"nearly broken"~\cite{Koberle}.

Finally, we remark that the Lagrangian (\ref{418}) with
added flavor degrees of freedom

\begin{equation}
{\cal L}=i\bar{\psi}\not\!\partial\psi +{1\over 2}
\lambda_{0}^{2}[(\bar{\psi}\psi )^{2}-
(\bar{\psi}\gamma^{5}\tau_{i}\psi)
(\bar{\psi}\gamma^{5}\tau_{i}\psi)]
\label{435}
\end{equation}

\noindent is a popular candidate
to study the low energy phenomenology of the
light mesons. In the above, the $\tau_{i}$ are taken as the generators
of $SU(2)$ for the two flavor model, and as the generators of $SU(3)$
for the three flavor model. The invariant correlators now carry four
labels

\begin{equation}
\sigma^{ij}_{\alpha\beta}(x,y)=\bar{\psi}^{ai}_{\alpha}(x)
\psi^{aj}_{\beta}(y)
\label{436}
\end{equation}

\noindent where $i,j$ are flavor indices.
It is not hard to construct the leading configuration

\begin{equation}
\sigma^{ij}_{\alpha\beta}(x,y)=
\delta^{ij}\int {d^{2}k\over (2\pi )^{2}}
e^{-ik(x-y)}\big[ {-i\over \not\! k-m_{r}}\big]_{\alpha\beta}
\label{437}
\end{equation}

\noindent where the mass
$m_{r}^{2}=\lambda^{4}(\sigma^{2}+\tilde{\sigma}^{2})$
satisfies the gap equation

\begin{equation}
{1\over 4i\lambda_{0}^{2}}=\int {d^{2}k\over (2\pi )^{2}}
{1\over k^{2}-2\lambda_{0}^{2}m_{r}^{2}}
\label{438}
\end{equation}

The effective field theory propagator is
easily written in terms of this leading
configuration

\begin{eqnarray}
\nonumber
A_{\mu\nu\mu'\nu'}^{iji'j'}&&(p_{1},p_{2},p_{3}p_{4})=
-\delta(p_{3}-p_{2})\delta(p_{1}-p_{4})
{\sigma^{0}}^{ij'}_{\mu\nu'}(p_{1})
{\sigma^{0}}^{i'j}_{\mu'\nu}(p_{3})\\
&&+{i\lambda_{0}^{2}\over (2\pi )^{2}}\delta (p_{1}+p_{3}-p_{2}-p_{4})
\sum_{M,N={\bf 1}\otimes{\bf 1},i\gamma^{5}\otimes\tau_{i}}
{[{\sigma^{0}}^{T}(p_{1})M{\sigma^{0}}^{T}(p_{2})]_{\nu\mu}^{ji}
[{\sigma^{0}}^{T}(p_{3})N{\sigma^{0}}^{T}(p_{4})]_{\nu'\mu'}^{j'i'}
\over 1+i\lambda_{0}^{2}\int {d^{2}k\over (2\pi )^{2}}
Trtr[{\sigma^{0}}^{T}M
{\sigma^{0}}^{T}(k-p_{1}+p_{2})N]}
\label{439}
\end{eqnarray}

\noindent where $Tr$ denotes a trace in Dirac space, $tr$ denotes a
trace in flavor space and $M,N$ are now direct products
of matrices belonging to Dirac space with matrices belonging
to flavor space.
\footnote{There is a slight abuse of notation in (\ref{439}) - 
please refer to (90-92) for clarification.}
Notice that in this case a triplet of massless pseudoscalar bosons
appears in the spectrum. These are usually interpreted as the
$\pi^{+}$, the $\pi^{-}$ and the $\pi^{0}$ particles.

In the context of this
phenomenological model, our effective field theory is nothing
but the theory of the mesons built from quark anti-quark
pairs. The couplings of the various
meson-meson interactions are proportional
to N (the number of quarks) raised to some negative (integer or
half integer) power. 

\subsection{Homogeneous Bethe Salpeter Equation.}

We have shown how to compute the Feynman rules and propagator for our
effective field theory and have verified their correctness perturbatively,
and non perturbativley to first nontrivial order in $1\over N$.
However before any scattering amplitudes can be
computed, we have to supply
a set of asymptotic states. $S$ matrix elements are then
taken using these asymptotic states, as usual.

In any field theory, only the quadratic term in the action provides
a harmonic oscillator and thus a spectrum consistent with a set of free
particles. Thus, it is the quadratic term that codes the asymptotic
states of the theory, as solutions of the corresponding homogeneous
equation. This should always correspond to the large $N$ approximation
to the homogenous Bethe Salpeter equation.

For example, the effective field theory wave functions
for the Gross Neveu model satisfy

\begin{equation}
i(\not\! p_{2} -m)_{\rho'\alpha}
\eta_{\rho\alpha}(p_{3},p_{2})
(\not\! p_{3} -m)_{\rho\alpha'}
+{\lambda_{0}^{2}\over 4\pi^{2}}\int dp_{1}dp_{4}
\delta (p_{1}+p_{3}-p_{2}-p_{4})\eta_{\alpha\alpha}(p_{4},p_{1})
\delta_{\alpha'\rho'}=0.
\label{wfunc}
\end{equation}

It is possible to obtain a particular solution to this equation. Making
the ansatz

\begin{equation}
\eta_{\rho\alpha}(p_{3},p_{2})=
\bar{\eta}_{\rho\alpha}(p_{3},p_{3}-p_{2}),
\label{soln}
\end{equation}

\noindent we find that $\bar{\eta}$ satisfies the equation

\begin{equation}
(\not\! p_{2} -m)_{\rho'\alpha}
\bar{\eta}_{\rho\alpha}(p_{2},p_{3}-p_{2})
(\not\! p_{3} -m)_{\rho\alpha'}=
i{\lambda_{0}^{2}\over 4\pi^{2}}\int dp_{1}
\bar{\eta}_{\alpha\alpha}(p_{1},p_{3}-p_{2})
\delta_{\alpha'\rho'}.
\label{wfunc2}
\end{equation}

Rewriting this last equation as

\begin{equation}
\bar{\eta}_{\mu\nu}(p_{2},p_{3}-p_{2})=
\big[{i\over \not\! p_{2} -m)}
{i\over\not\! p_{3} -m}\big]_{\mu\nu}
i{\lambda_{0}^{2}\over 4\pi^{2}}\int dp_{1}
\bar{\eta}_{\alpha\alpha}(p_{1},p_{3}-p_{2})
\label{wfunc3}
\end{equation}

\noindent taking the trace of both sides and integrating over
$p_{2}$ keeping $p_{3}-p_{2}=k$ fixed,
we are lead to the consistency condition

\begin{equation}
1=
i{\lambda_{0}^{2}\over 4\pi^{2}}\int dp_{2}
\big[{i\over \not\! p_{2} -m}
{i\over\not\! p_{2}+\not\! k -m}\big]_{\nu\nu}
\label{wfunc4}
\end{equation}

This condition requires $k^{2}=4m^{2}$ which,
since $k$ is the physical momentum transfer for the
particle antiparticle channel,
corresponds to the leading order
mass shell condition
for the fermion-antifermion bound state.

\section{Subleading ${1\over N}$ Corrections.}
The formalism which was developed in sections $II$ and $III$ can 
be used to obtain subleading corrections to any 
correlator of interest. For instance the ${1\over N}$ correction to
the propagator results from the following diagrams in the effective field
theory

\begin{equation}
\setlength{\unitlength}{1mm}
\begin{picture}(150,15)
\put(10,5){$<\delta\sigma\delta\sigma>=$}

\put(30,5){$\times$}
\put(40,5){$\times$}
\put(31.25,5.75){\line(1,0){10}}
\put(44,5){$+$}
\put(36.25,5.75){\circle*{2}}

\put(47,5){$\times$}
\put(57,5){$\times$}
\put(48.25,5.75){\line(1,0){10}}
\put(61,5){$+$}
\put(53.25,7.75){\circle{4}}

\put(64,5){$\times$}
\put(74,5){$\times$}
\put(65.25,5.75){\line(1,0){10}}
\put(78,5){$+$}
\put(70.25,5.75){\line(0,1){3}}
\put(70.25,10.75){\circle{4}}

\put(81,5){$\times$}
\put(91,5){$\times$}
\put(82.25,5.75){\line(1,0){10}}
\put(95,5){$+$}
\put(87.25,5.75){\line(0,1){3}}
\put(87.25,9.75){\circle*{2}}

\put(98,5){$\times$}
\put(108,5){$\times$}
\put(99.25,5.75){\line(1,0){3}}
\put(109.25,5.75){\line(-1,0){3}}
\put(104.25,5.75){\circle{4}}

\end{picture}
\label{66}
\end{equation}

\noindent with the Feynman rules given by 
$(95)$ in terms of the leading
configuration $\sigma_{0}$
and where the propagators in the diagrams refer to (114). It is 
essential that both the disconnected and connected pieces
of this propagator are included.
One observes that the contributions from
the subleading term of the Jacobian (proportional to
$L^{d}\delta^{d}(0)$) precisely cancel similar infinite contributions
generated in the other diagrams.  All the remaining expressions
then have an interpretation in terms of original Feynman diagrams.
In this precise sense the subleading term in the Jacobian provides a normal
ordering. Diagrammatically, we find:

\begin{equation}
\setlength{\unitlength}{1mm}
\begin{picture}(140,120)


\put(5.75,89.5){\line(1,0){12}}
\put(5.75,84.5){\line(1,0){12}}
\put(5.75,89.5){\line(-1,-1){6}}
\put(5.75,84.5){\line(-1,1){6}}

\put(11.75,99.5){\circle{2}} 
\put(11.75,97.5){\circle{2}} 
\put(11.5,95.75){$.$}
\put(11.5,95){$.$}
\put(11.5,94.25){$.$}
\put(11.75,92.5){\circle{2}} 
\put(11.75,90.5){\circle{2}} 
\put(11.75,106){\circle{11}}
\put(7.25,106){\circle{2}} 
\put(9.25,106){\circle{2}} 
\put(10.25,105.75){$...$} 
\put(14.25,106){\circle{2}} 
\put(16.25,106){\circle{2}} 


\put(25,89){$+$}

\put(40.75,84.5){\line(-1,1){7}}
\put(40.75,90.5){\line(-1,-1){7}}
\put(40.75,84.5){\line(1,0){15}}
\put(40.75,90.5){\line(1,0){15}}
\put(41.75,91.5){\circle{2}}
\put(49.75,91.5){\circle{2}} 
\put(42.95,93.2){\circle{2}} 
\put(48.55,93.2){\circle{2}} 
\put(44.2,94.2){$...$}

\put(64,100){\line(1,-1){18}}
\put(89,100){\line(-1,-1){18}}
\put(71.75,93.5){\circle{2}}
\put(73.75,93.5){\circle{2}}
\put(74.75,93.5){$...$}
\put(78.75,93.5){\circle{2}}
\put(80.75,93.5){\circle{2}}

\put(148,48.75){$+$}

\put(112,53){\line(1,-1){4}}
\put(112,45){\line(1,1){4}}
\put(117,49){\circle{2}} 
\put(119,49){\circle{2}} 
\put(120,48.75){$...$} 
\put(124,49){\circle{2}} 
\put(126,49){\circle{2}} 
\put(127,49){\line(1,2){2.75}}
\put(127,49){\line(1,-2){2.75}}
\put(129.75,54.5){\line(1,0){12}}
\put(129.75,43.5){\line(1,0){12}}

\put(135.75,64.5){\circle{2}} 
\put(135.75,62.5){\circle{2}} 
\put(135.5,60.75){$.$}
\put(135.5,60){$.$}
\put(135.5,59.25){$.$}
\put(135.75,57.5){\circle{2}} 
\put(135.75,55.5){\circle{2}} 
\put(135.75,71){\circle{11}}
\put(131.25,71){\circle{2}} 
\put(133.25,71){\circle{2}} 
\put(134.25,70.75){$...$} 
\put(138.25,71){\circle{2}} 
\put(140.25,71){\circle{2}} 

\put(62,89){$+$}

\put(96,93){\line(1,-1){4}}
\put(96,85){\line(1,1){4}}
\put(101,89){\circle{2}} 
\put(103,89){\circle{2}} 
\put(104,88.75){$...$} 
\put(108,89){\circle{2}} 
\put(110,89){\circle{2}} 
\put(111,89){\line(1,2){2.75}}
\put(111,89){\line(1,-2){2.75}}
\put(113.75,83.5){\line(1,0){20}}
\put(113.75,94.5){\line(1,0){20}}

\put(119.75,95.5){\circle{2}} 
\put(127.75,95.5){\circle{2}} 
\put(120.95,97.2){\circle{2}} 
\put(126.55,97.2){\circle{2}} 
\put(122.2,98.2){$...$}

\put(89,89){$+$}
\put(136,89){$+$}

\put(5,53){\line(1,-1){4}}
\put(5,45){\line(1,1){4}}
\put(10,49){\circle{2}} 
\put(12,49){\circle{2}} 
\put(13,48.75){$...$} 
\put(17,49){\circle{2}} 
\put(19,49){\circle{2}} 
\put(22,49){\circle{4}}
\put(25,49){\circle{2}} 
\put(27,49){\circle{2}} 
\put(28,48.75){$...$} 
\put(32,49){\circle{2}} 
\put(34,49){\circle{2}} 
\put(35,49){\line(1,-1){4}}
\put(35,49){\line(1,1){4}}
\put(22,61){\circle{2}} 
\put(22,59){\circle{2}} 
\put(21.75,57.25){$.$}
\put(21.75,56.5){$.$}
\put(21.75,55.75){$.$}
\put(22,54){\circle{2}} 
\put(22,52){\circle{2}} 
\put(22,67.5){\circle{11}}
\put(17.5,67.5){\circle{2}} 
\put(19.5,67.5){\circle{2}} 
\put(20.5,67.25){$...$} 
\put(24.5,67.5){\circle{2}} 
\put(26.5,67.5){\circle{2}} 

\put(56,48.75){$+$}

\put(62,53){\line(1,-1){4}}
\put(62,45){\line(1,1){4}}
\put(67,49){\circle{2}} 
\put(69,49){\circle{2}} 
\put(70,48.75){$...$} 
\put(74,49){\circle{2}} 
\put(76,49){\circle{2}} 
\put(83,49){\circle{11}}
\put(89.5,49){\circle{2}} 
\put(91.5,49){\circle{2}} 
\put(92.5,48.75){$...$} 
\put(96.5,49){\circle{2}} 
\put(98.5,49){\circle{2}} 
\put(99.5,49){\line(1,1){4}}
\put(99.5,49){\line(1,-1){4}}

\put(79,54.5){\circle{2}} 
\put(87,54.5){\circle{2}} 
\put(80.2,56.2){\circle{2}} 
\put(85.8,56.2){\circle{2}} 
\put(81.45,57.2){$...$}

\put(106.25,48.75){$+$}

\put(5,33){\line(1,-1){4}}
\put(5,25){\line(1,1){4}}
\put(10,29){\circle{2}} 
\put(12,29){\circle{2}} 
\put(13,28.75){$...$} 
\put(17,29){\circle{2}} 
\put(19,29){\circle{2}} 
\put(25.5,29){\circle{11}}
\put(25.5,33.5){\circle{2}} 
\put(25.5,31.5){\circle{2}} 
\put(25.25,29.75){$.$}
\put(25.25,29){$.$}
\put(25.25,28.25){$.$}
\put(25.5,26.5){\circle{2}} 
\put(25.5,24.5){\circle{2}} 
\put(32,29){\circle{2}} 
\put(34,29){\circle{2}} 
\put(35,28.75){$...$} 
\put(39,29){\circle{2}} 
\put(41,29){\circle{2}} 
\put(42,29){\line(1,1){4}}
\put(42,29){\line(1,-1){4}}

\put(56,28.75){$+$}

\put(62,33){\line(1,-1){4}}
\put(62,25){\line(1,1){4}}
\put(67,29){\circle{2}} 
\put(69,29){\circle{2}} 
\put(70,28.75){$...$} 
\put(74,29){\circle{2}} 
\put(76,29){\circle{2}} 
\put(77,29){\line(1,2){2.75}}
\put(77,29){\line(1,-2){2.75}}
\put(79.75,23.5){\line(1,0){12}}
\put(79.75,34.5){\line(1,0){12}}

\put(82.5,33.5){\circle{2}} 
\put(82.5,31.5){\circle{2}} 
\put(82.25,29.75){$.$}
\put(82.25,29){$.$}
\put(82.25,28.25){$.$}
\put(82.5,26.5){\circle{2}} 
\put(82.5,24.5){\circle{2}} 

\put(99.5,28.75){$+$} 

\put(128.5,28.75){$+$}

\put(5,13){\line(1,-1){4}}
\put(5,5){\line(1,1){4}}
\put(10,9){\circle{2}} 
\put(12,9){\circle{2}} 
\put(13,8.75){$...$} 
\put(17,9){\circle{2}} 
\put(19,9){\circle{2}} 
\put(22,9){\circle{4}}
\put(23.5,11.5){\circle{2}} 
\put(23.5,6.5){\circle{2}} 
\put(25.5,11.5){\circle{2}} 
\put(25.5,6.5){\circle{2}} 
\put(26.5,11.25){$...$} 
\put(26.5,6.25){$...$} 
\put(30.5,11.5){\circle{2}} 
\put(30.5,6.5){\circle{2}} 
\put(32.5,11.5){\circle{2}} 
\put(32.5,6.5){\circle{2}} 
\put(34,9){\circle{4}}
\put(37,9){\circle{2}} 
\put(39,9){\circle{2}} 
\put(40,8.75){$...$} 
\put(44,9){\circle{2}} 
\put(46,9){\circle{2}} 
\put(47,9){\line(1,1){4}}
\put(47,9){\line(1,-1){4}}

\put(56,8.75){$+$}

\put(62,13){\line(1,-1){4}}
\put(62,5){\line(1,1){4}}
\put(67,9){\circle{2}} 
\put(69,9){\circle{2}} 
\put(70,8.75){$...$} 
\put(74,9){\circle{2}} 
\put(76,9){\circle{2}} 
\put(79,9){\circle{4}}
\put(80.5,11.5){\circle{2}} 
\put(80.5,6.5){\circle{2}} 
\put(82.5,11.5){\circle{2}} 
\put(82.5,6.5){\circle{2}} 
\put(83.5,11.25){$...$} 
\put(83.5,6.25){$...$} 
\put(87.5,11.5){\circle{2}} 
\put(87.5,6.5){\circle{2}} 
\put(89.5,11.5){\circle{2}} 
\put(89.5,6.5){\circle{2}} 
\put(90.5,11.5){\line(1,0){4}}
\put(90.5,6.5){\line(1,0){4}}
\put(90.5,6.5){\line(0,1){5}}

\put(105.5,31.5){\line(1,0){4}}
\put(105.5,26.5){\line(1,0){4}}
\put(109.5,26.5){\line(0,1){5}}

\put(110.5,31.5){\circle{2}} 
\put(110.5,26.5){\circle{2}} 
\put(112.5,31.5){\circle{2}} 
\put(112.5,26.5){\circle{2}} 
\put(113.5,31.25){$...$} 
\put(113.5,26.25){$...$} 
\put(117.5,31.5){\circle{2}} 
\put(117.5,26.5){\circle{2}} 
\put(119.5,31.5){\circle{2}} 
\put(119.5,26.5){\circle{2}} 
\put(120.5,31.5){\line(1,0){4}}
\put(120.5,26.5){\line(1,0){4}}
\put(120.5,26.5){\line(0,1){5}}

\end{picture}
\label{1000}
\end{equation}

In the above, we have only included basic skeletons; permutations
of these basic skeletons need to be included. For example, the
first diagram appearing would be repeated with another diagram in
which the second fermion is dressed. 

It is well known that the Gross Neveu model is
exactly integrable with no particle production 
(although it has a rich spectrum~\cite{DHN}) and
the exact $S$ matrices are known~\cite{Zam}.
For two particle scattering the $S$ matrix element is given by

\begin{equation}
^{out}<P_{b}(\tilde{p}_{1})P_{d}(\tilde{p}_{2})|
P_{a}(p_{1})P_{c}(p_{2})>^{in}=
_{ac}S_{bd}(\theta )\delta(\tilde{p}_{1}^{1}-p_{1}^{1})
\delta(\tilde{p}_{2}^{1}-p_{2}^{1})-
_{ac}S_{db}(\theta )\delta(\tilde{p}_{1}^{1}-p_{2}^{1})
\delta(\tilde{p}_{2}^{1}-p_{1}^{1}).
\label{scattering}
\end{equation}

\noindent where

\begin{equation}
_{ac}S_{bd}(\theta )=\sigma_{2}(\theta,2N)\delta_{ab}\delta_{cd}
+\sigma_{3}(\theta,2N)\delta_{ad}\delta_{bc},
\label{aaa}
\end{equation}

In the above equations 
$\theta=\theta_{1}-\theta_{2}$ 
where $\theta_{1}$ and $\theta_{2}$ are the rapidity variables
of the incoming particles.
The above $U(N)$ symmetric
two particle $S$ matrix is consistent 
with the underlying $O(2N)$ symmetry of the Gross
Neveu model~\cite{berg2}. $\sigma_{2}$ and
$\sigma_{3}$ are the standard symbols used to describe the exact $S$ matrix
with $O(N)$ symmetry.

It turns out that the simplest way to verify the 
$1\over N$ correction to the 
propagator is to consider the contributions of the diagrams in
(\ref{1000}) to forward two particle scattering corresponding to 
time flowing from top to bottom. The corresponding $S$ matrix is then

\begin{equation}
_{ab}S_{ab}(forward)=N^{2}(\sigma_{2}+{1\over N}\sigma_{3}).
\label{this}
\end{equation}

By looking at the diagrams in (\ref{1000}) we see 
that the first two diagrams can not contribute to forward
scattering. The third diagram is clearly the leading ($O({1\over N})$)
contribution to $\sigma_{3}$. The remaining diagrams  should
sum up to the $1\over N^{2}$ contribution to $\sigma_{2}$. They are
precisely the diagrams considered by Berg et. al.~\cite{ScatAmp} who
have indeed confirmed that this is the case.

\section{Conclusions}

In this paper, we have used colorless bilocal fields to study the large $N$
limit of both fermionic and bosonic vector models. By requiring that the
Schwinger Dyson equations in terms of the original variables agree with the
equations derived directly in terms of the invariant variables led to a
functional differential equation for the Jacobian. The equation was solved
exactly, leading to an exact effective action. This effective action 
was then shown to reproduce the familiar perturbative expansion for the two
and four point functions. In particular, in the case of fermionic vector
models, the effective action correctly accounts for the  Fermi statistics. 
The theory was then studied non-perturbatively. The stationary points of the
effective action provide the usual large $N$ gap equations. The homogeneous
equation associated with the quadratic (in the bilocals) action is simply a
two particle Bethe Salpeter equation. Finally, the leading correction in 
$1\over N$ was shown to be in agreement with the exact $S$ matrix of the model.

There are a number of interesting questions which can now be asked. Firstly, it
is clear that the above invariant variables are classicaly commuting functions.
This change of variables therefore provides a bosonization valid in an arbitrary
number of dimensions. It is interesting to ask if a link can be drawn to more
conventional bosonization schemes. Also, this bosonization may prove to be a
powerful tool for analyzing many body condensed matter systems. Secondly, it
is well known that the large $N$ limit of the Gross-Neveu model posseses a
number of interesting configurations, corresponding to many particle bound
states. The present formalism is particularly well suited to studying two
particle bound states. It would thus be interesting to determine the Feynman
rules of the effective theory in a two particle bound state background. One
would expect that these rules may be simpler than the rules obtained in this
paper. Finally, the possibility of obtaining an equal time approach for
fermionic vector models remains of great interest.\par
\newpage

\appendix{Appendix A: Diagrams to integrals.}

In this appendix, explicit expressions are supplied for the fermionic
diagrams appearing in the text (see equation(\ref{84})):

\begin{equation}
\setlength{\unitlength}{1mm}
\begin{picture}(150,14)
\put(5,5){\line(1,0){12}}
\put(19,5){$={i\over \not\, p-m}$}
\end{picture}
\label{500}
\end{equation}

\begin{equation}
\setlength{\unitlength}{1mm}
\begin{picture}(150,14)
\put(5,5){\line(1,0){12}}
\put(12,7){\circle{4}}
\put(12,5.5){\circle*{0.5}}
\put(12,4.5){\circle*{0.5}}
\put(19,5){$=-i\sum_{M={\bf 1},i\gamma^{5}}
{i\over \not\, p-m}
\int {d^{d}k\over (2\pi )^{d}}Tr[M{i\over \not\, k-m}]
M{i\over \not\, p-m}$}
\end{picture}
\label{501}
\end{equation}

\begin{equation}
\setlength{\unitlength}{1mm}
\begin{picture}(150,14)
\put(5,5){\line(1,0){12}}
\put(8,7){\circle{4}}
\put(8,5.5){\circle*{0.5}}
\put(8,4.5){\circle*{0.5}}
\put(14,7){\circle{4}}
\put(14,5.5){\circle*{0.5}}
\put(14,4.5){\circle*{0.5}}
\put(19,5){$=i^{2}\sum_{M,N={\bf 1},i\gamma^{5}}
{i\over \not p-m}
\int {d^{d}k\over (2\pi )^{d}}Tr[M{i\over \not k-m}]
M{i\over \not p-m}                          
\int {d^{d}k'\over (2\pi )^{d}}Tr[N{i\over \not k'-m}]
N{i\over \not p-m}$}
\end{picture}
\label{502}
\end{equation}

\begin{equation}
\setlength{\unitlength}{1mm}
\begin{picture}(150,14)
\put(5,5){\line(1,0){12}}
\put(11,7){\circle{4}}
\put(11,5.5){\circle*{0.5}}
\put(11,4.5){\circle*{0.5}}
\put(11,11){\circle{4}}
\put(11,9.5){\circle*{0.5}}
\put(11,8.5){\circle*{0.5}}
\put(21,5){$=i^{2}\sum_{M,N={\bf 1},i\gamma^{5}}
{i\over \not p-m}
\int {d^{d}k\over (2\pi )^{d}}
Tr[M{i\over \not k-m}N{i\over \not k-m}]
\int {d^{d}k'\over (2\pi )^{d}}Tr[N{i\over \not k'-m}]
M{i\over \not p-m}$}
\end{picture}
\label{503}
\end{equation}

\begin{equation}
\setlength{\unitlength}{1mm}
\begin{picture}(150,14)
\put(5,5){\line(1,0){12}}
\put(11,7){\circle{4}}
\put(9,5.5){\circle*{0.5}}
\put(13,5.5){\circle*{0.5}}
\put(19,5){$=i\sum_{M={\bf 1},i\gamma^{5}}
{i\over \not\, p-m}
\int {d^{d}k\over (2\pi )^{d}}M{i\over \not\, k-m}
M{i\over \not\, p-m}$}
\end{picture}
\label{504}
\end{equation}

\begin{equation}
\setlength{\unitlength}{1mm}
\begin{picture}(150,14)
\put(5,5){\line(1,0){12}}     
\put(8,7){\circle{4}}          
\put(8,5.5){\circle*{0.5}}      
\put(8,4.5){\circle*{0.5}}       
\put(14,7){\circle{4}}             
\put(16,5.5){\circle*{0.5}}          
\put(12,5.5){\circle*{0.5}}            
\put(19,5){$=-i^{2}\sum_{M,N={\bf 1},i\gamma^{5}}
{i\over \not p-m}
\int {d^{d}k\over (2\pi )^{d}}Tr[M{i\over \not k-m}]
M{i\over \not p-m}                          
\int {d^{d}k'\over (2\pi )^{d}}N{i\over \not k'-m}
N{i\over \not p-m}$}
\end{picture}
\label{505}
\end{equation}

\begin{equation}
\setlength{\unitlength}{1mm}
\begin{picture}(150,14)
\put(5,5){\line(1,0){12}} 
\put(8,7){\circle{4}}      
\put(10,5.5){\circle*{0.5}}  
\put(6,5.5){\circle*{0.5}}
\put(14,7){\circle{4}}
\put(14,5.5){\circle*{0.5}}
\put(14,4.5){\circle*{0.5}}
\put(19,5){$=-i^{2}\sum_{M,N={\bf 1},i\gamma^{5}}
{i\over \not p-m}
\int {d^{d}k\over (2\pi )^{d}}M{i\over \not k-m}
M{i\over \not p-m}                          
\int {d^{d}k'\over (2\pi )^{d}}Tr[N{i\over \not k'-m}]
N{i\over \not p-m}$}
\end{picture}
\label{506}
\end{equation}

\begin{equation}
\setlength{\unitlength}{1mm}
\begin{picture}(150,14)
\put(5,5){\line(1,0){12}}      
\put(11,7){\circle{4}}           
\put(13,5.5){\circle*{0.5}}        
\put(9,5.5){\circle*{0.5}}          
\put(11,11){\circle{4}}                
\put(11,9.5){\circle*{0.5}}             
\put(11,8.5){\circle*{0.5}}               
\put(21,5){$=-i^{2}\sum_{M,N={\bf 1},i\gamma^{5}}
{i\over \not p-m}
\int {d^{d}k\over (2\pi )^{d}}
M{i\over \not k-m}N{i\over \not k-m}
\int {d^{d}k'\over (2\pi )^{d}}Tr[N{i\over \not k'-m}]
M{i\over \not p-m}$}
\end{picture}
\label{507}
\end{equation}

\begin{equation}
\setlength{\unitlength}{1mm}
\begin{picture}(150,14)
\put(5,5){\line(1,0){12}}
\put(11,7){\circle{4}}
\put(11,5.5){\circle*{0.5}}
\put(11,4.5){\circle*{0.5}}
\put(11,11){\circle{4}}
\put(9,9){\circle*{0.5}}
\put(13,9){\circle*{0.5}}
\put(21,5){$=-i^{2}\sum_{M,N={\bf 1},i\gamma^{5}}
{i\over \not p-m}
\int {d^{d}k\over (2\pi )^{d}}
\int {d^{d}k'\over (2\pi )^{d}}
Tr[M{i\over \not k-m}N{i\over \not k'-m}
N{i\over \not k -m}]
M{i\over \not p-m}$}
\end{picture}
\label{508}
\end{equation}

\begin{equation}
\setlength{\unitlength}{1mm}
\begin{picture}(150,14)
\put(5,6){\line(1,0){12}}
\put(11,6){\circle{4}}
\put(9.5,5.5){\circle*{0.5}}
\put(8.5,6.5){\circle*{0.5}}
\put(13.5,6.5){\circle*{0.5}}
\put(12.5,5.5){\circle*{0.5}}
\put(19,5){$=-i^{2}\sum_{M,N={\bf 1},i\gamma^{5}} 
{i\over \not p-m}M
\int{d^{d}k\over (2\pi )^{d}}
{d^{d}k'\over (2\pi )^{d}}
{i\over \not k-m}N
Tr[{i\over \not k+\not k'-\not p-m}N
{i\over \not k'-m}M]
{i\over \not p-m}$}
\end{picture}
\label{509}
\end{equation}

\end{document}